%% file: ms.tex
\documentclass{emulateapj}
\slugcomment{{\sc Accepted for publication in ApJ} }
\shortauthors{Toft et al.}
\shorttitle{ A correlation between size and star formation activity at $z\sim 2.5$}
\begin{document}

\title{HST and Spitzer imaging of red and blue galaxies at $z\sim2.5$: A correlation between size and star formation activity from compact quiescent galaxies to extended star forming galaxies \altaffilmark{1}}

\author{S. Toft\altaffilmark{2,}\altaffilmark{3}, P. van Dokkum\altaffilmark{2}, M. Franx\altaffilmark{4}, I. Labbe\altaffilmark{5,}\altaffilmark{6}, N.M. F\"{o}rster Schreiber\altaffilmark{7}, S. Wuyts\altaffilmark{4},  T. Webb\altaffilmark{8}, G. Rudnick\altaffilmark{9}, A. Zirm\altaffilmark{10}, M. Kriek\altaffilmark{4}, P. van der Werf\altaffilmark{4}, J.P. Blakeslee\altaffilmark{11}, G. Illingworth\altaffilmark{12}, H.-W. Rix\altaffilmark{13},  C. Papovich\altaffilmark{14},  A. Moorwood\altaffilmark{3}  }

\altaffiltext{1}
{Based on observations made with the NASA/ESA \emph {Hubble Space Telescope}, which is operated by the Association of Universities for Research in Astronomy, Inc, under NASA contract NAS5-26555, observations made with the Spitzer Space Telescope, which is operated by the Jet Propulsion Laboratory, California Institute of Technology under NASA contract 1407 and observations collected at the European Southern Observatory, Paranal, Chile (ESO Program 164.O-0612)}

\altaffiltext{2}
{Department of Astronomy, Yale University, New Haven, CT 06520-8101, USA.}

\altaffiltext{3}
{European Southern Observatory, Karl-Schwarzschild-Str. 2, D-85748 Garching bei M\"{u}nchen, Germany, email: stoft@eso.org.}

\altaffiltext{4}
{Sterrewacht Leiden, Leiden University, Postbus 9513, NL-2300 RA Leiden, Netherlands.}

\altaffiltext{5}
{Carnegie Observatories, 813 Santa Barbara Street, Pasadena, CA 91101, USA}

\altaffiltext{6}
{Carnegie Fellow}

\altaffiltext{7}
{Max-Planck-Institut f\"{u}r extraterrestrische Physik, Giessenbachstrasse, D-85748 Garching, Germany.}

\altaffiltext{8}
{Department of Physics, McGill University, 3600 rue University, Montr\'{e}al, Qu\'{e}bec, Canada}

\altaffiltext{9}
{NOAO, 950 Cherry Avenue, Tuescon, AZ 85719, USA}

\altaffiltext{10}
{Center for Astrophysical Sciences, Johns Hopkins University, 3400 North Charles Street, Baltimore, MD 21218, USA}

\altaffiltext{11}
{Dept. of Physics \& Astronomy, Washington State University, Pullman, WA 99164-2814, USA}

\altaffiltext{12}
{University of California, UCO/Lick Observatory, Santa Cruz, CA 95064,
USA}

\altaffiltext{13}
{Max-Planck-Institut f\"{u}r Astronomie, K\"{o}nigstuhl 17, 69117 Heidelberg, Germany}

\altaffiltext{14}
{Steward Observatory, University of Arizona 933 North Cherry Avenue, Tucson, AZ 85721, USA }

\begin{abstract}   
We present HST NICMOS+ACS and Spitzer IRAC+MIPS observations of 41 galaxies at $2<z<3.5$ in the FIRES MS1054 field with red and blue rest-frame optical colors. 
About half of the galaxies are very compact (effective radii $r_e< 1$ kpc) at rest-frame optical wavelengths, the others are extended ($1< r_{e}< 10$ kpc). 
 For reference, 1 kpc corresponds to $0\farcs12$ at $z=2.5$ in the adopted cosmology. 
We separate actively star forming galaxies from quiescent galaxies by modeling their rest-frame UV-NIR SEDs.
The star forming galaxies span the full range of sizes, while the quiescent galaxies all have $r_e<2$kpc.
In the redshift range where MIPS $24\micron$ imaging is a sensitive probe of re-radiated dust emission ($z<2.5$), the  $24\micron$ fluxes confirm that the light of the small quiescent galaxies is dominated by old stars, rather than dust-enshrouded star formation or AGN activity.  
The inferred surface mass densities and velocity dispersions for the quiescent galaxies are very high compared to those in local galaxies.
The galaxies follow a Kormendy relation (between surface brightness and size) with approximately the same slope as locally, but shifted to brighter surface brightnesses, consistent with a mean stellar formation redshift of $z_f\sim5$.
This paper demonstrates a direct relation between star formation activity and size at $z\sim2.5$, and the existence of a significant population of massive, extremely dense, old stellar systems without readily identifiable counterparts in the local universe.
\end{abstract}

\keywords{galaxies: formation --- galaxies: evolution --- galaxies: high redshift --- galaxies: fundamental parameters --- galaxies: structure --- infrared: galaxies}

\section{INTRODUCTION}
In the local universe, galaxy structure correlates with physical properties such as star formation rate and history, mass, environment, gas and dust content, metallicity etc. This is likely a consequence of the fact that star formation activity scales with environment and structure (bulge to disk ratio) scales with mass \citep{kauffmann2004}. High density environments are dominated by early type galaxies which are homogeneously old, red, quiescent and with low dust contents, while lower density environments are dominated by late type galaxies which on average are bluer, less massive and show a range of ages, dust contents, star formation histories, metallicities etc. \citep[e.g.][]{goto04, baldry2006}. Observations suggests that these correlations exist out to at least $z\sim 1.4$, where clusters are still dominated by uniformly large, old, red early type galaxies with little ongoing star formation and dust, although the fraction of blue late type galaxies in clusters increase with redshift \citep[e.g][]{blakeslee2003, toft2004,bell2004,mcintosh2005, lidman2004, mullis2005, cooper2007, cassata2007}.   
Understanding the epoch and nature of the onset of these relations are key issues for understanding galaxy formation, and it is essential to push simultaneous studies of structure and star formation properties of galaxies to higher redshifts. 

Until recently, most work on high redshift galaxies was done using rest-frame UV selected Lyman break galaxies, which were thought to be representative of the high redshift galaxy population. An HST study comparing the optical (rest-frame UV) WFPC2 and NIR (rest-frame optical) NICMOS structure of Lyman break galaxies in the Hubble Deep Field North (HDFN) showed that these were similar \citep{Dickinson2000}. 
This result discouraged further NICMOS observations of high redshift galaxies in the following years, since NICMOS observations are much less efficient than optical WFPC2 and ACS observations. In a number of studies \citep{Giavalisco1996, Lowenthal1997, Dickinson2000, Ferguson2004, bouwens2004} it was assumed that rest-frame UV structural properties (such as Hubble-type and size) were representative of the rest-frame optical structural properties (which traces the bulk of the stellar mass distribution, and is related to the dynamical state of the galaxies). 
The similarity of the rest-frame optical and UV structural properties of high redshift galaxies in the HDFN was later confirmed in a more detailed study, which calculated an internal UV-optical color dispersion, and showed that this quantity was small for galaxies at $z\sim2.3$ \citep{papovich2005}.

The Faint InfraRed Extragalactic Survey \citep[FIRES,][]{Labbe2003, Franx2003}  discovered a population of distant ($2<z<3.5$) red galaxies (DRGs) using the simple selection criterion $J-Ks>2.3$, which proved very efficient for locating massive evolved galaxies and massive galaxies with significant amounts of dust and ongoing star formation at these redshifts. Since then a number of studies have employed this selection technique to study different aspects of the properties of DRGs, and their relation to other high redshift populations of galaxies.
Based on optical-NIR SED fits it has been demonstrated that DRGs on average have higher stellar masses, higher mass-to-light ratios, more dust and higher star formation rates \citep{forster-schreiber2004} than the rest-frame UV selected Lyman Break Galaxies (LBGs).  
DRGs dominate in stellar mass limited ($M>10^{11}M_{\sun}$) samples \citep{vandokkum2006} and the bright end of the rest-frame optical luminosity function \citep{marchesini2006} over galaxies with $J-K_s<2.3$ in the same redshift range (for convenience we label these galaxies ``distant blue galaxies'' (DBGs) in the following) and  they  cluster more strongly \citep{quadri2007}. DRGs also dominate the stellar mass density in luminosity selected samples \citep{rudnick2003,rudnick2006}, over DBGs. 
DBGs share many properties with LBGs (they are younger, less massive and less obscured than DRGs), but there is not a one to one correspondence. The majority of of LBGs have ($J-K<2.3$), but a not insignificant subset have
redder $J-K$ colors \citep[see][]{quadri2006musyc}, and DBGs are not necessarily LBGs, some of them do not meet the $U_nGR$ color criterion \citep{steidel1996} or are too faint in the $R$ band \citep{quadri2006musyc, vandokkum2006}. 
Recently, the addition of mid infrared (MIR) observations from the Spitzer Space telescope has allowed a better separation between DRGs which are red due to old passively evolving stellar populations, and DRGs which are red due to dust obscured star formation \citep{labbe2005, papovich2006, webb2006, reddy2006}.

In the GOODS-S field about $50\%$ of the DRGs (in the redshift range $1.5<z<2.5$ where MIPS is sensitive to star formation, see Sec.\ref{mipsvssize}) were detected with MIPS at $24\micron$ \citep{papovich2006},  and a similar fraction of DRGs were detected in the GOODS-N field \citep{reddy2006}. The MIPS data on these fields was not very deep (only sensitive to $SFR>100M_{\sun} yr^{-1}$ at $z=2$), and it was argued, based on other diagnostics of star formation, that $<10\%$ of the DRGs in the GOODS-S field have rest-frame UV-optical-IR colors consistent with quiescent stellar populations \citep{papovich2006}.
In the Extended HDFS, MIPS $24\micron$ observations detected $65\%$ of the DRGs in the same redshift range, indicating that this fraction is dusty actively star forming galaxies \citep{webb2006}.
Based on their non-detection by MIPS at $24\micron$, the remaining DRGs ($35-50\%$) are presumably red due to more quiescent stellar populations.
This picture is broadly consistent with NIR spectroscopy of a sample of 20 bright ($K<19.6$) DRGs in the redshift range $2<z<2.6$ which found that a surprisingly large fraction ($45\%$) of these had no emission lines, and continuum breaks consistent with evolved stellar populations \citep{kriek2006}.

Interestingly, DRGs show a difference between their rest-frame UV and optical structural properties: In a study based on the Hubble Ultra Deep Field NICMOS \citep{thompson2005} and ACS data \citep{Beckwith2006}, it was demonstrated that the structure of DRGs depends strongly on wavelength \citep{Toft2005}. The six DRGs in this field were shown to have regular centrally symmetric rest-frame optical structure, while the rest-frame UV structure, as observed with ACS were much more irregular, and thus not representative of the underlying stellar mass distribution of the galaxies. This study however did not include rest-frame UV selected galaxies to test whether this was also the case for them.

Recently, the existence of a population of extremely dense massive galaxies at $z> 1.2$  have been suggested by a number of studies. In the UDF, 2/6 DRGs with $2<z<3.5$ were found to be unresolved by NICMOS $H_{F160W}$ band observations \citep{Toft2005}, corresponding to rest-frame optical physical sizes of $r_e \lesssim 0.8 kpc$, a factor of $\sim5$ smaller than galaxies of similar mass in the local universe. In another study of the UDF, 4/7  spectroscopically confirmed massive ($M > 10^{11}M_{\sun}$) early-type galaxies with $1.4<z<2.5$ were found to have similarly small physical sizes $r_e \lesssim 0.8 kpc$, measured in the ACS $z_{F850LP}$ band \citep{daddi2005}.
In a study at lower redshifts ($1.2<z<1.7$), 10 massive ($\sim 5\times10^{11}M_{\sun})$ galaxies in the MUNICS survey  were shown to be a factor of $\sim4$ smaller than local galaxies of similar mass.
 In a NICMOS $H_{F160W}$ band study of the HDFS, 5/14 DRGs with $2 \lesssim z \lesssim 3.5$ were found to have $r_e\lesssim 1$ kpc \citep{zirm2007}. This study also suggested the existence of correlation between star formation activity and size at $z\sim 2.5$.  The rest-frame optical-NIR SEDs of the compact  ($r_e<1 kpc$) DRGs were better fitted by single stellar population models without dust, than constantly star forming model with dust, while the opposite was true for the more extended ($r_e > 1 kpc$) DRGs.
The main uncertainties of the above studies are that they are based on very small samples, and the possibility that AGN may be dominating the light in the compact galaxies, and affect their size measurements, mass and star formation estimates. 
Interrestingly, recent dynamical studies of sub-millimeter galaxies (SMGs) at $z \sim 2\rm{-}3$ also imply high matter densities, from large dynamical masses $\sim 10^{11} M_{\odot}$  within radii $\lesssim 2 kpc$ \citep{tacconi2006,bouche2007}
In this paper we present a study based on deep NICMOS and ACS observations of a sample of 27 DRGs and a sample of 14 DBGs in the redshift range $2<z<3.5$ in the FIRES MS1054-03 field \citep{Forster-Schreiber2006}.
In addition to an analysis of how the structure of DRGs and DBGs compare to each other and depend on wavelength, we combine the ground based optical-NIR FIRES photometric catalog with deep MIR Spitzer Space telescope IRAC photometry to construct rest-frame UV-NIR SEDs for the galaxies, from which we constrain properties of the stellar populations (star formation history, stellar mass, dust content, etc.) and investigate how these correlate with structural properties derived from the HST data.  We also add deep MIPS $24\micron$ imaging to the analysis, and demonstrate how size correlates with $24\micron$ flux.

Using this much larger sample of galaxies (and spanning a larger range of colors and masses) than previous studies, we demonstrate that there is indeed a correlation between star formation activity and size $z\sim2.5$, ranging from extremely compact quiescent galaxies to extended star forming galaxies. Furthermore, using diagnostics based on X-ray and  MIPS $24\micron$ observations we independently demonstrate that the small sizes of the quiescent galaxies are not caused by AGN activity. 

The paper is organized as follows:  
In Sec.~\ref{sec.data} we describe sample selection and the data, including details of the HST and Spitzer data reduction and photometry, and the construction of optical-MIR photometric catalogs and SED fits.
In Sec.\ref{morphologies} we present the HST images of the DRGs and DBGs and describe their basic structural properties. In this section we also quantify the degree to which the structure depend on wavelength and the degree to which this depends on $J-K$ color.  
In Sec.\ref{profiles} we fit the 2D surface brightness distributions of the galaxies with Sersic profiles, and determine their best fitting Sersic n-parameters and effective radii $r_e$.
In Sec.\ref{sec.sizes} we compare the distributions of $r_e$ and n-parameters for DRGs and DBGs, and for star forming and quiescent galaxies.
In Sec \ref{mipsvssize} we present $24\micron$ MIPS imaging of the sample, and show how the $24\micron$ flux correlates with the size of the galaxies.
In Sec.\ref{agn} we use different diagnostics to search for the presence of AGN in the sample
In Sec.\ref{correlations} we explore the dataset for correlations between size and properties of the stellar populations.  
We derive relations between mass and size,
mass and surface mass density, mass and velocity dispersion, and size and surface brightness (the Kormendy relation), and compare to local relations to constrain a possible evolution.
In Sec.\ref{models} we compare the results to model predictions and speculate on an evolutionary link between the observed galaxies, and local galaxies. 
In Sec.\ref{caveats} we discuss the uncertainties and possible caveats of the analysis, and in Sec.\ref{summary} we summarize the results. 

We assume a standard cosmology ($h_0=0.7, \Omega_{M}=0.3, \Omega_{\Lambda}=0.7$) throughout the paper, and all quoted magnitudes are in the Vega system. When deriving masses we assume a Salpeter initial mass function. 

\section{DATA: SELECTION, PHOTOMETRY AND PHOTOMETRIC REDSHIFTS}
\label{sec.data}
We used the FIRES data to construct a sample of galaxies with photometric redshifts $z_{phot} \ga 2$ in the MS1054-03 field. Photometry and photometric redshifts were adopted from \citet{Forster-Schreiber2006}.  We distinguish between distant red galaxies (DRG), selected by the criterion $J-Ks>2.3$ which efficiently isolates rest-frame optically red galaxies at $z>2$, and distant blue galaxies (DBGs), those that do not meet the $J-Ks>2.3$ criterion but have $z_{phot}>2$.   

\subsection{NICMOS Observations}
\label{nicmos}
\begin{figure*}
\plotone{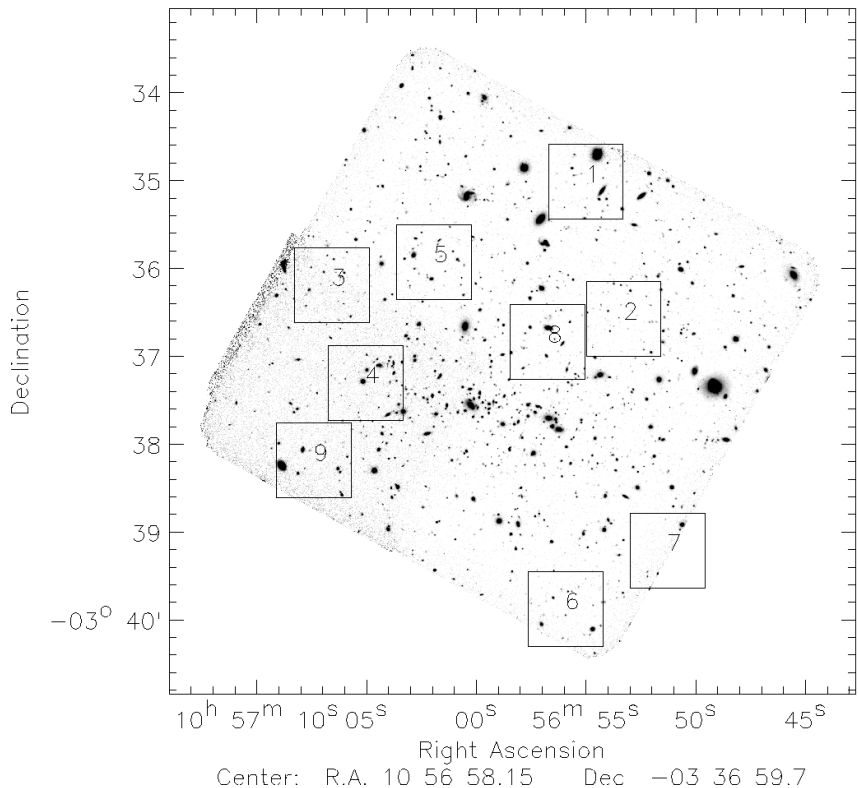}
\caption{ 
\label{nic_layout}
Layout of 9 NICMOS pointings on the FIRES K-band image \citep{Forster-Schreiber2006}. North is up, east is to the left.  
}
\end{figure*}

We obtained nine NICMOS pointings in the F160W band ($H_{F160W}$) with camera 3 ($52\arcsec \times 52\arcsec$, 0\farcs203/pix).  The pointings were chosen to include as many DRGs as possible and as a secondary criteria, to include as many $z>2$ DBGs as possible. Using this scheme we end up with a sample of 27 DRGs and 14 DBGs with NICMOS observations. At the time of observation, the NIC3 camera was out of focus, which result in a slightly broader PSF, than when the camera is in focus.
In Fig.\ref{nic_layout} we show the layout of the NICMOS pointings on the FIRES K-band image \citep{Forster-Schreiber2006}. 

Each pointing was split up into six sub-pixel dithered exposures to allow for a better sampling of the point spread function and cosmic ray rejection.
The six pointings were reduced and combined using the following recipe:
a first order estimate of the sky, a pedestal signal (which is present in all NICMOS images) was subtracted from the individual frames using the IRAF task pedsky. Next, the background subtracted images were cross correlated to determine their shifts with respect to each other, using the iraf task crossdrizz. Using these shifts the background subtracted images were drizzled to a common coordinate system, with a finer pixel scale (with pixfrac$=0.65$), using the iraf task drizzle. The drizzled images were then median combined and the median combined image de-drizzled back to the pixel scale and coordinate system of the individual images, using the iraf task blot. The ``blotted'' images were then compared to the original images to create cosmic-ray maps, and cross-correlated to refine the shifts.
 Finally the cosmic ray corrected, background subtracted images were drizzled using the refined shifts. The final combined images have a pixel scale $0.1\arcsec$, and an exposure time of 1.5 hours. 
The combined drizzled images showed considerable residual large scale ($\sim 25\arcsec$) features. These were successfully fitted and subtracted from the combined drizzled images, using background maps generated by {\sc SExtractor} with options set to mask out objects and save background maps estimated using a fine mesh fitting.  The masking thresholds passed to {\sc sextractor} was chosen very conservatively to make sure that size estimates of faint galaxies and the outer parts of low surface brightness galaxies were not affected. 

\subsection{ACS Observations}
\label{ACS}
The MS1054 field was observed with ACS for 2 hours in the F606W-band ($V_{F606W}$) and for 5 hours in each of the F775W ($i_{F775W}$) and F850LP ($z_{F850LP}$) bands \citep{blakeslee2006}. Most of the DRGs in the sample are very faint in the rest-frame UV emission sampled by ACS, so we stacked the three ACS images to increase the signal to noise. The summed $V_{F606W}+i_{F775W}+z_{F850LP}$ image has a total exposure time of 12 hours, and a pixel scale of 0.05\arcsec.
Details of the reduction of the ACS data can be found in \cite{blakeslee2006}.

\subsection{Spitzer IRAC Observations}
\label{sec.irac}
We observed the FIRES MS1054 field with the Mid-Infrared IRAC instrument on-board the Spitzer Space Telescope with a series of dithered observations resulting in combined $5.5\arcmin \times 5.5\arcmin$ images with exposure time of $\sim$3.2 hours in each of the 3.6, 4.5, 5.8 and 8.0 $\micron$ bands.  
The data reduction and photometry was carried out in a similar way as in \cite{Labbe2006} and \cite{Wuyts2007} and details of the data reduction and photometry method are described in Labb\'{e} et al (in prep).
Here we briefly describe the data reduction and photometry. We started with the basic calibrated data (BCD) as provided by the Spitzer Science center. We applied a series of procedures to reject cosmic rays and remove artifacts such as column pull-down, muxbleed, and the ``first frame effect'' \citep{Hora2004}. Finally, the frames were registered to a $2\times2$ block averaged ($0\farcs2396$ pixel scale) version of the existing ISAAC $5.5 \arcmin \times 5.3 \arcmin$ K-band mosaic \citep{Forster-Schreiber2006}. Unfortunately, even though the size of the FIRES K-band mosaic is similar to the size of the deep IRAC pointings, the overlap is only $\sim70\%$, due to different centering and rotation of the two data sets.
 
The main challenge in doing IRAC photometry is a proper treatment of source confusion.
Information on the position and shape of the K-band detected objects were used to fit and subtract the fluxes of neighboring sources.
Each K-band source is isolated using the {\sc SExtractor} ``segmentation'' map and convolved individually with a special kernal to match the IRAC PSF. Next all convolved sources were fitted to the IRAC image, leaving only the fluxes as free parameters. Subsequently the best fitting fluxes of neighboring sources were subtracted to remove contamination. 
 This correction is relatively small for most of the galaxies. The contribution to the total $3.6\micron$-band flux from neighboring sources is $<10\%$ for 21 of the DRGs, $10-20\%$ for 5 of the DRGs and $\sim40\%$ for 1 DRG (with similar numbers in the other bands).
We measure fluxes on the cleaned images, within $3\arcsec$ apertures, and apply to each source an aperture correction to scale the fluxes by the ``color'' apertures defined in the FIRES catalog (the correction is the ratio of the K-band flux in the color aperture to the K-band flux within the $3\arcsec$ aperture).
The end product is a photometric catalog with consistent photometry from optical to MIR wavelengths with 12 filters ($UBVV_{F606W}I_{F814W}JHK$+IRAC), from which we extract the SEDs of the $z>2$ galaxies in the sample. Note that the $V_{F606W}$ and $I_{F814W}$ filters in the catalog are from earlier WFPC2 observations, not the ACS observations discussed in Sec.\ref{ACS}.     

\subsection{Spitzer MIPS Observations}
The MS1054 field was observed with the Multi-band Imaging Photometer for the {\em Spitzer Space Telescope} \citep[MIPS; ][]{rieke2004} as part of the GTO Lensing Cluster Survey \citep[e.g.][]{egami2007}.  The images were obtained using the $24\micron$ channel ($\lambda_{c}=23.7 \micron$; $\Delta\lambda = 4.7 \mu$m) which uses a 128$\times$128 BIB Si:As array with a pixel scale of 2\farcs55 pixel$^{-1}$.  The data were reduced and combined with the Data Analysis Tool (DAT) developed by the MIPS instrument team \citep{gordon2005}, and a few additional processing steps were applied as described in \citet{egami2006}. The resulting mosaic has a pixel size of $0\farcs5$. The total integration time is 3600~s over most of the 5\arcmin$\times$5\arcmin\ field centered on
the cluster, and is as much as 4800~s in a $\sim30\arcsec$-wide strip
crossing the cluster center because of the way multiple data sets were
taken. The 5$\sigma$ detection limit is $\sim$50 $\mu$Jy for a point source.

We applied the same scripts to the mosaic to de-blend the light from confused sources and perform aperture photometry, as was applied to the IRAC data (see Sec.\ref{sec.irac}).
We derive fluxes in $6\arcsec$ apertures after cleaning for flux from neighboring sources, and applied aperture corrections to derive total fluxes. In Tab.\ref{modeltable} we list the $24\micron$ MIPS flux of the galaxies in the sample.
 A larger fraction of the DRGs than the DBGs are detected, both in the full redshift range ($55\%$ vs $15\%$), and at redshifts $z<2.5$ ($70\%$ vs $25\%$). This difference is most likely caused by higher average star formation rate in the DRGs compared to in the DBGs (see Sec.\ref{mipsvssize})
  
\subsection{SED Modeling}
\label{sedmodel}
We fitted the rest-frame UV-NIR SEDs of the galaxies using the BC2003 stellar population synthesis models \citep{bc2003}, assuming solar metallicity, a Salpeter IMF between 0.1 and 100 $M_{\odot}$ and a Calzetti extinction law.  For the purpose of determining the stellar mass, dust content, age and star formation rate of the galaxies we fitted models with three different star formation histories: a single burst model with no dust (SSP), an exponentially decaying star formation rate model, with a characteristic timescale of 300Myr (tau) and a constantly star forming model (CSF), both with variable amounts of dust. For each of the star formation histories, we constrained the time elapsed since the onset of star formation to $>50 Myr$, to avoid unrealistically young ages.

To parameterize the star formation history of the galaxies, we divided the sample into ``quiescent'' and ``star forming'' galaxies, based on whether the SSP or CSF model provided a better fit to the SED. 
We chose to use only these two extreme star formation histories to limit the number of free parameters, and keep the division as simple as possible.
Ideally one would make the division based on specific star formation rate (sSFR), but this quantity is not well constrained from SED fits.
We emphasize that galaxies are generally expected to lie between these extremes. Our goal is to divide the population in two broad groups, based on their star formation rate, not to determine accurate star formation histories of individual galaxies. 
We note that the distinguishment between the SSP and CSF models is only used to divide the sample in to star forming and quiescent galaxies. For the purpose of deriving parameters of the SED fits, we also include the tau model, to avoid biases which could be introduced by fitting only the two extreme star formation histories.

 In table \ref{modeltable} we list  properties of the DRG and DBG samples.
 $26\%/15\%$ of the DRGs/DBGs  are classified as quiescent, and the remaining  $74\%/85\%$ are classified as star forming. 
We note that DRG 25 is labeled star forming, but is equally well fit by the quiescent model (the $\chi^2$ of the CSF model fit is only marginally better than the $\chi^2$ of the SSP model fit).
 The derived stellar  masses of the DRGs are in the range $[1.8\rm{-}55.4]\times10^{10}M_{\odot}$, with a median $\left < M \right>_{DRG}= 16.9 (\pm 4.2) \times 10^{10} M_{\odot}$. The masses of the DBG are in the range $[0.4\rm{-}14.0]\times10^{10}M_{\odot}$, with a median $\left < M \right>_{DRG}= 2.5 (\pm 0.7) \times 10^{10} M_{\odot}$, a factor of $\sim 7$ lower than for DRGs. 
This difference in mass is in agreement with previous studies  and is mainly a consequence of the DRGs having larger mass to light ratios ($M/L$) than the DBGs, due to older ages and a higher dust content \citep[see][]{forster-schreiber2004, labbe2005}. 
While degeneracies between estimates of e.g. dust content and stellar age from restframe UV-NIR SEDs fits can be substantial, stellar masses are in general quite robust \citep{vandokkum2006}, and more precise than stellar mass estimates derived from rest frame K-band emission alone, due to the large observed variations in the rest frame $M/L_K$ of galaxies at $z\simeq 2.5$ \citep[see][]{labbe2005}.

Further general details on the method, ingredients of the SED modeling, and its strengths and limitations are described in \cite{forster-schreiber2004} and \cite{Wuyts2007}. A detailed analysis of the SED modeling for the present dataset will be presented in Forster-Schreiber et al (in prep.), but the main uncertainties and caveats related to the mass estimation are discussed in Sec.\ref{caveats}.

\begin{figure*}
\plotone{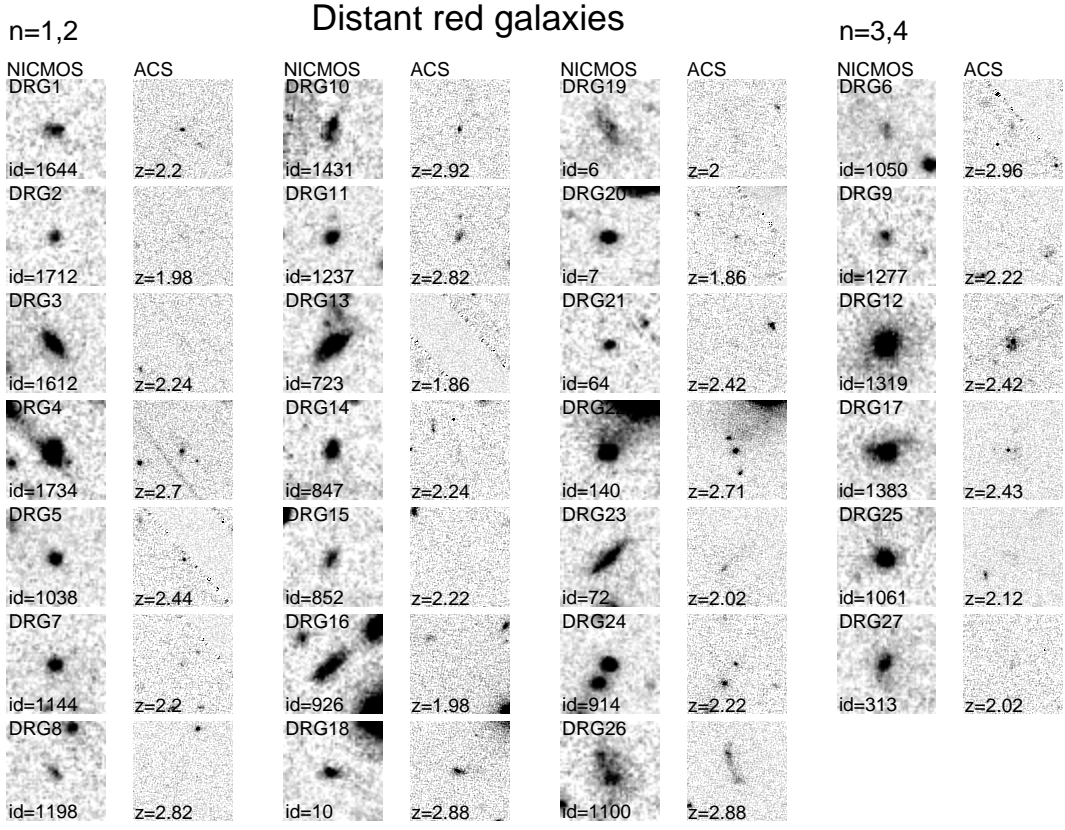}
\caption{ 
\label{drg}
Images of the 27 DRGs in the sample. NICMOS $H_{F160W}$-band (1.5h) and stacked  $V_{F606W}+i_{F775W}+z_{F850LP}$ (12h) ACS images are shown side by side. The 21 galaxies in the first three columns have exponential disk-like surface brightness profiles, while the 6 galaxies in the fourth column have  ``de Vaucouleurs-like'' surface brightness profiles. Four galaxies (DRG 5, 20, 21 and 24) are better fitted by exponential disk-like surface brightness profiles, but are formally unresolved. The size of the images is $5\arcsec \times 5\arcsec$. }
\end{figure*}

\begin{figure*}
\plotone{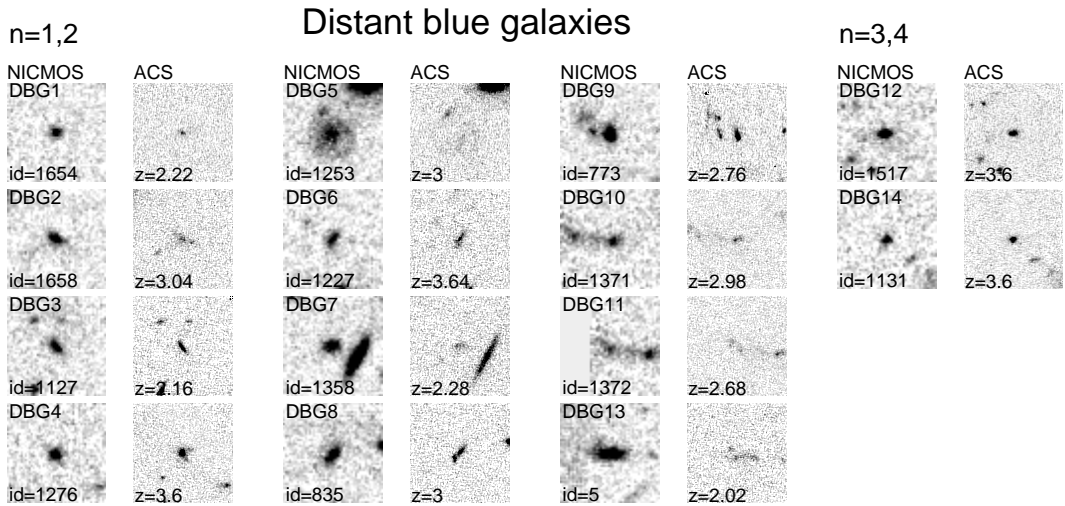}
\vspace{-4cm}
\caption{Images of the 14 DBGs in the sample. NICMOS $H_{F160W}$-band (1.5h) and stacked  $V_{F606W}+i_{F775W}+z_{F850LP}$ (12h) ACS images are shown side by side. The 12 galaxies in the first three columns have exponential disk-like surface brightness profiles, while the 2 galaxies in the fourth column are better fitted by  ``de Vaucouleurs-like'' surface brightness profiles (but are formally unresolved). The size of the images is $5\arcsec \times 5\arcsec$.
\label{dbg}
}
\end{figure*}

\section{GALAXY STRUCTURE}
\label{morphologies}
In Fig.\ref{drg} we show a gallery of the DRGs in the sample as they appear in the NICMOS images and in the stacked ACS image. 
The DRGs show a range of structural properties in the NICMOS images. About half of them are compact, the others are extended.  
At first glance, one might classify some of these galaxies as ellipticals, irregulars or mergers, but we abstain from such a classification here,
and derive quantitative parameters, concentrating on half-light radius
and Sersic index (see Sec.\ref{profiles}). 
Visual inspection of the NICMOS images suggest that about $37\%$ of the DRGs have close faint (DRG 4, 11, 22) or bright (DRG 13, 24) companions or assymetries (10, 14, 17, 26, 27) which indicate they could be in the process of minor or major mergers. Unfortunately it is not possible to derive reliable photometric redshifts for any of the possible companions due to their faint magnitudes and/or small angular seperations.
We can not establish whether they are real physical associations.
The difference between the appearance of the DRGs at rest-frame optical and rest-frame UV wavelengths is striking: While most of the DRGs are detected at high S/N in the NICMOS images, more than half of them are only marginally detected, or not detected at all in the ACS images. Some of the most extreme examples are DRGs 3, 13 and 16 which are among the brightest galaxies in the NICMOS images, but are undetected in the ACS images.  
The DRGs that are detected in the ACS images in some cases have quite different structures in the NICMOS and ACS images. The most pronounced examples are the DRG 17, 23 and 26, which are among the largest galaxies in the NICMOS images. In the NICMOS image DRG 23 appears like a nice symmetric edge on disk galaxy, but in the ACS image the disk is almost invisible, and the emission is dominated by a an off-center ``blob'' at the edge of the disk. In the NICMOS image, DRG 26 appears like a face-on disk galaxy dominated by a bright central bulge, and a large lower surface brightness diffuse disk with  a few off center blobs of emission, but in the ACS image the galaxy appears like a ``chain'' galaxy, dominated by a blob of emission in one end of the chain rather than by emission from the central bulge. This galaxy is detected in relatively shallow Chandra X-ray observations and thus hosts an AGN (see Sec.\ref{agn}). In the NICMOS image, DRG 17 appears to have a bright central bulge, and two asymmetrical arms, which may be either spiral arms or tidal tails induced by a recent merger. In the ACS image all that is detected is a double central point-source, and a bit of diffuse emission. This galaxy is has been directly detected with SCUBA, indicating that it hosts very intense star formation \citep{knudsen2005}.

In Fig.\ref{dbg} we show a similar gallery of the DBGs in the sample. 
Comparing this figure to Fig.\ref{drg}, the main differences are 1) that 
a larger fraction of DBGs ($57\%$) have faint (DBGs 3, 5) or bright (DBG 10, 11) nearby companions or show assymetries (DBGs 2, 4, 6, 8, 9, 13) that may be contributed to minor or major recent or ongoing merging (note that companion of DBG 7 has a lower $z_{phot}$, and that DBG 10 and 11 which are two components of the same major merger are counted as two seperate galaxies), 2) all the DBGs are detected in the ACS image, and 3) their structure (with the exception of DBG5) are similar in the NICMOS and ACS images.

\begin{figure}
\plotone{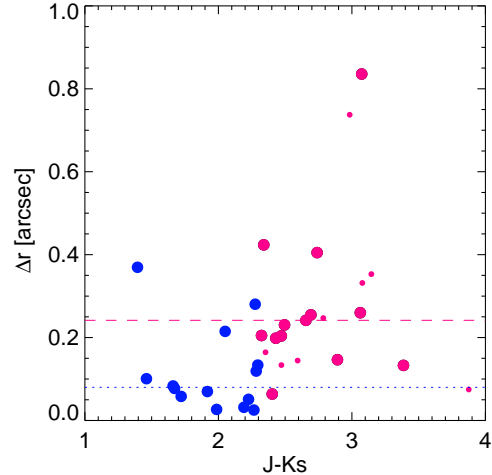}
\caption{Offset $\Delta r$ between the luminosity weighted central pixel position of the light distribution in the ACS and NICMOS images, as a function of their  $J-Ks$ color. The luminosity weighed central pixel is calculated as $(x_c,y_c)=(\sum_{x,y}{x\cdot i(x,y)}/\sum_{x,y}i(x,y),\sum_{x,y}{y\cdot i(x,y)})/\sum_{x,y}i(x,y)$, where $i(x,y)$ is the intensity in the pixel with coordinates (x,y), and x and y run over all pixel values with intensities $>3 \sigma$ in a $3\arcsec\times 3\arcsec$ box centered on the galaxy  . 
The red symbols represent DRGs, the blue symbols represent DBGs. Large symbols are detected at high significance in the ACS images, small symbols are marginally detected. The dashed red line is the median offset of the DRGs  $\left<\Delta r\right >_{DRG}=0.24\pm0.03$ arcsec, the dotted blue line is the median offset of the DBGs  $\left<\Delta r\right >_{DBG}=0.08\pm0.03$ arcsec .
\label{jkdr}
}
\end{figure}

To illustrate the last point, we calculated the distance
between the luminosity weighted central pixel position of the light distribution  in the ACS and NICMOS images, a measure of the difference between the ACS and NICMOS structure. In Fig.\ref{jkdr} this quantity is shown to correlate with J-K color. The median shift for the DRGs is larger (by a factor of $\sim3$) than for the DBGs. According to the Mann-Whitney test there is a  $99.6\%$ chance of the mean shifts being different.  We did not find evidence for a correlation between $\Delta r$ and $H_{F160W}$ band flux. 
Note that the fact that the galaxies are very faint in the very deep ACS imaging informs us about their colors, but not about potential morphological 
differences. In fact, it makes quantitative  statements very hard. Hence, we use only a very simple indicator, the centroid difference, to study the difference between rest-frame UV and optical. A proper analysis would compare sizes, asymmetry, etc. \citep[see e.g.,][]{papovich2005}.

The relatively relaxed centrally symmetric NICMOS structure and faint and more disturbed ACS structure of the DRGs are consistent with composite stellar populations, where relatively dusty, evolved red populations dominate the rest-frame optical structure (and stellar masses), and off-center blobs of ongoing star-formation in some cases dominate the rest-frame UV structure.
This demonstrates that high resolution rest-frame optical observations are essential for deriving structural information and indirectly, information on the dynamical state of massive red galaxies at $z>2$ (e.g. sizes, Hubble types etc).
The DBGs, on the other hand, have very similar rest-frame optical and UV structure, suggesting that they are dominated by uniformly young populations of stars with little dust.
This suggests that the conclusion about the size distribution and structural properties of LBGs based on their observed optical properties \citep[e.g.][]{Ferguson2004, bouwens2004} are probably ``correct'' since their rest-frame UV structure is likely to be representative of their rest-frame optical structure.

Note that the prominent differences between the restframe UV and optical structure of the red galaxies in some cases may introduce systematic errors in photometric studies, based on aperture photometry from lower spatial resolution data, as flux in bluer waveband apertures may originate in different physical regions of the galaxies than the flux in the redder waveband apertures, making the assumption that a monotonic star formation history can be fitted to the integrated flux of the galaxy invalid.           

\section{PROFILE FITS AND SIZES}
\label{profiles}
We fitted the 2D surface brightness distribution of the galaxies in the sample, using {\sc galfit} \citep{peng2002} and a S\`ersic profile: $I(r)=I(0)exp\left[b_n(r/r_e)^{1/n}\right]$, which has been shown to be a good representation of the surface brightness of a range of galaxy types from ellipticals (n=4) to disks (n=1). The effective radius  $r_e$, is enclosing half the light of the model and $b_n$ is a normalization constant. {\sc galfit} returns the effective radius along the semi-major axis of the model, $a_e$. This can be converted to the ``circularized'' effective radius $r_e=a_e\sqrt{1-\epsilon}$, where $\epsilon$ is the projected ellipticity of the galaxy model.    
To limit the degree of freedom in the fits, we fixed n at the values 1, 2, 3 and 4 and identified the best fitting model by comparing the $\chi^2$ of the four fits. The best fitting values of $n$ and the corresponding best fitting values of $r_e$ are listed in Table \ref{sizetable}.

{\sc Galfit} fits models which are convolved with the PSF (estimated directly from stars in the field) to the data, and is thus in principle capable of measuring sizes smaller than the size of the PSF. There are four stars in our NICMOS pointings which are sufficiently isolated and bright to be used as PSF stars. 
We repeated the analysis using each star in turn to estimate the effect of the choice of PSF star on the distribution of the derived sizes and radial shape parameters. No systematic dependence on the choice of PSF was found. The $1\sigma$ mean uncertainty of the derived size of a given galaxy due to the choice of PSF is $\sim 14\%$. The derived $n$-parameters are not significantly affected by the choice of PSF. We also estimated the effect of the additional background subtraction (described in Sec.~\ref{nicmos}), and the size of the fitting region on the derived sizes. No systematics offsets are found for these and the mean uncertainty due to these effects are of the order of $1\%$. Added in quadrature the total error due to additional background subtraction, and choice of PSF and the size of the fitting region is less than $15\%$.  
We estimated the minimum reliable size derived by galfit by applying the size fitting analysis to the four PSF stars, using the other three stars as PSF models. The mean size of ``bona fide'' unresolved objects is in this way estimated to be $\left < r_e \right >_{stars}=0\farcs030\pm0\farcs015$. Galaxies with derived sizes larger than $0\farcs075$ are thus resolved at the $3\sigma$ level in the NICMOS data.  
Four of the DRGs and 2 of the DBGs are unresolved in the NICMOS images, corresponding to a fraction of about 15$\%$ for both samples. In the following, we adopt $0\farcs075$ as an upper limit for the size of the unresolved objects.
We note that the galaxies that are unresolved in the NICMOS images have ACS sizes consistent with the upper limits derived from the NICMOS images.

\begin{figure}
\plotone{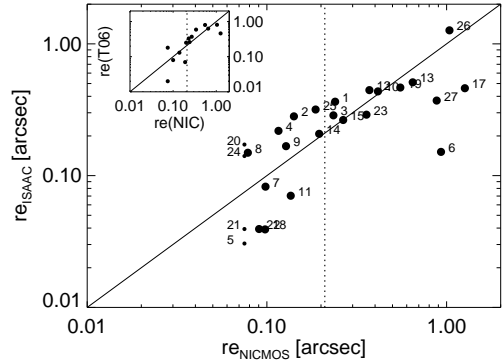}
\caption{ 
\label{compare_sizes}
Comparison of sizes derived from the NICMOS $H_{F160W}$ and the FIRES ISAAC $H$ band images. The small symbols represent galaxies which are unresolved in the NICMOS data (plotted at the NICMOS resolution limit). The dotted line represent the resolution limit of the ISAAC data. The numbers corresponds to the IDs of the DRGs. 
To illustrate the limitation of the ISAAC data for measuring the smallest sizes we plot galaxies at their {\em measured} ISAAC sizes, rather than correcting these to the resolution limit of the ISAAC data. 
 There is a relatively good agreement at sizes $re>0.2$ arcsec where the galaxies are resolved in the ISAAC images, at smaller sizes, there is a large scatter. The three galaxies that are significantly larger in the NICMOS images (DRG 6, 17 and 27) have low surface brightness components that are not detected by ISAAC.
The inset compares the sizes derived here from the NICMOS data ($re_{NIC}$) with the sizes derived in \citet{trujillo2006} from the FIRES data ($re_{T06}$), for the 14 galaxies which are in both data sets.
}
\end{figure}

As an independent check we fitted the surface brightness distributions and derived sizes for the 27 DRGs in the deep FIRES ISAAC H-band observation \citep{forster-schreiber2004}. This data is deeper but has a courser resolution (FWHM$\sim0.5\arcsec$). In Fig.\ref{compare_sizes} we compare the sizes derived for the DRGs in the ground based and space based data. There is no significant systematic offset between the sizes measured in the two data sets (median $\left< re_{ISAAC}/re_{NICMOS}\right >=1.06 \pm 0.17$), but there is a relatively large scatter, especially at sizes $re_{isaac}\lesssim0.2\arcsec$ where the galaxies are unresolved in the ISAAC data. 
Three galaxies which are resolved in both data sets, have significantly larger sizes in the NICMOS data, than in the ISAAC data. Visual inspection of the NICMOS images (see Fig.\ref{drg}) reveals that this is probably due to low surface brightness components which are not properly detected in the ISAAC data. A subset (14) of the DRGs studied here are also part of the sample of \citet{trujillo2006}, who derived sizes from the FIRES data in a very similar way as is done here for the ISAAC data. In the inset of Fig.\ref{compare_sizes} we compare the sizes of these galaxies derived by \citet{trujillo2006} from the ISAAC data with the sizes derived here from the NICMOS observations. Again, a relatively large scatter is evident, especially at sizes $re<0.2 \arcsec$, where the sizes determined from the ground based data is uncertain by at least a factor of 2. 

In addition to the systematic uncertainties in determining the angular sizes of the galaxies, there are uncertainties in the conversion of the angular sizes into physical sizes, due to uncertainties in the photometric redshifts. These are discussed in Sec.\ref{caveats}

\section{SIZE AND n-DISTRIBUTIONS} 
\label{sec.sizes}
In Tab.\ref{sizetable} we list the best fitting $n$ and $r_e$ parameters.
The galaxies show a broad size distribution, ranging from compact unresolved galaxies with $r_e<1$kpc, to large galaxies with $r_e>10$kpc. Most of the galaxies are best fitted by exponential disks. In the following we compare the size and n-distributions of different subsamples.

\begin{figure}
\epsscale{1.15}
\plottwo{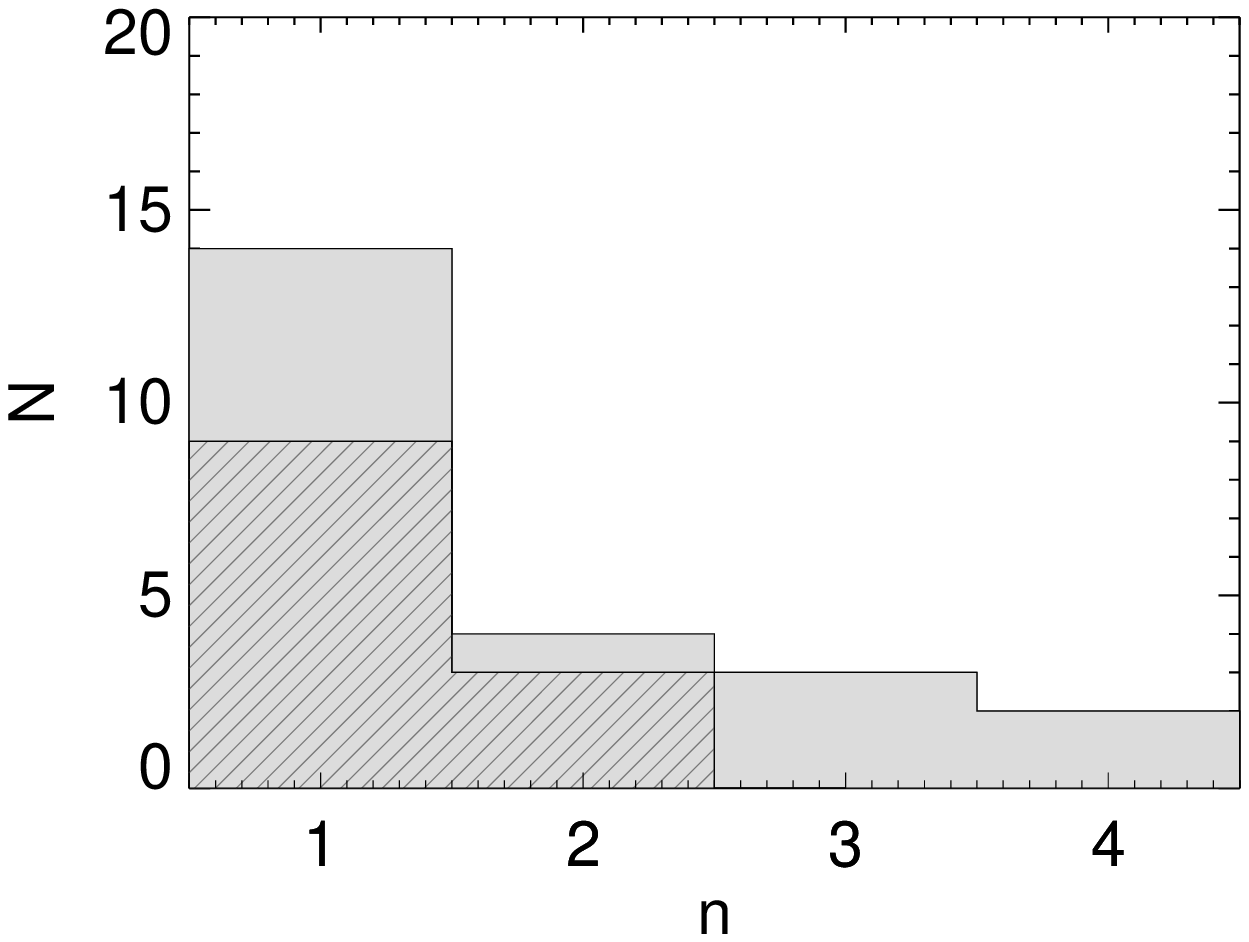}{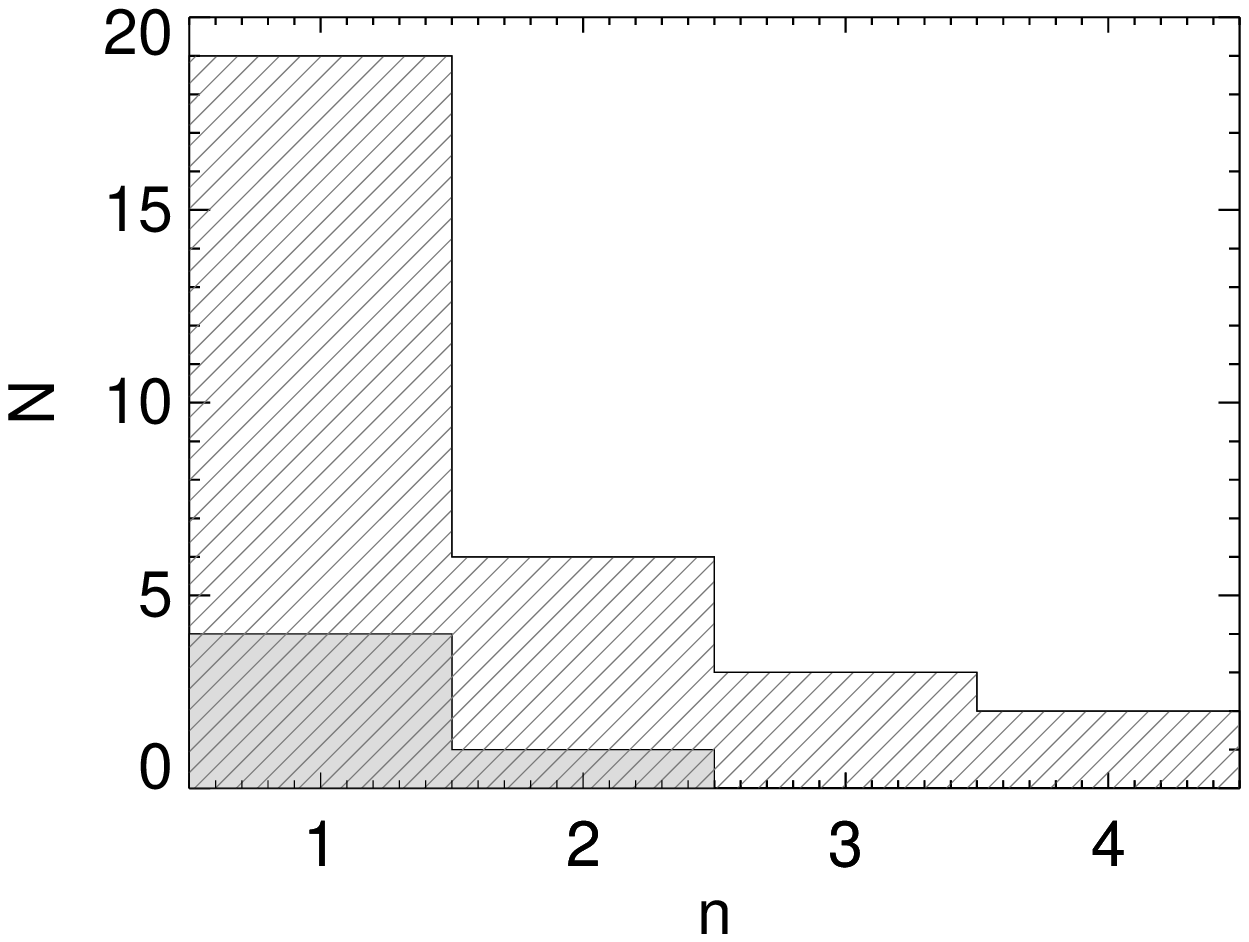}
\caption{Left: Distribution of Sersic index for resolved DRGs (shaded) and DBGs (hatched). Right:  Distribution of Sersic index for resolved quiescent (shaded) and star forming (hatched) galaxies (DRGs+DBGs).  
 In the left plot $80\%$ of the DRGs are better fitted by  exponential disks than $r^{1/4}$ law surface brightness profiles. In the right plot $83\%$ of the star forming galaxies are better fitted by  exponential disks than $r^{1/4}$ law surface brightness profiles
\label{n-distributions}
}
\end{figure}

In Fig.\ref{n-distributions} we plot the n-distribution of DRGs and DBGs. 
Both samples are dominated by galaxies with exponential disk profiles: $80\%$ of the resolved DRGs are best fitted by ``exponential disk-like'' surface brightness profiles ($n=1,2$) while $20\%$ are better fitted by ``de Vaucouleurs-like'' surface brightness profiles. All the resolved DBGs are best fitted by exponential disk-like laws. 
The KS-test yields a $21\%$ chance that the n-values were drawn from the same parent distribution. 
The distribution of radial shape parameters of the DRGs are similar to what was found in the UDF where 5 out of 6 DRGs where found to be best fit by exponential disk profiles \citep{Toft2005}.

\begin{figure}
\epsscale{1.15}
\plottwo{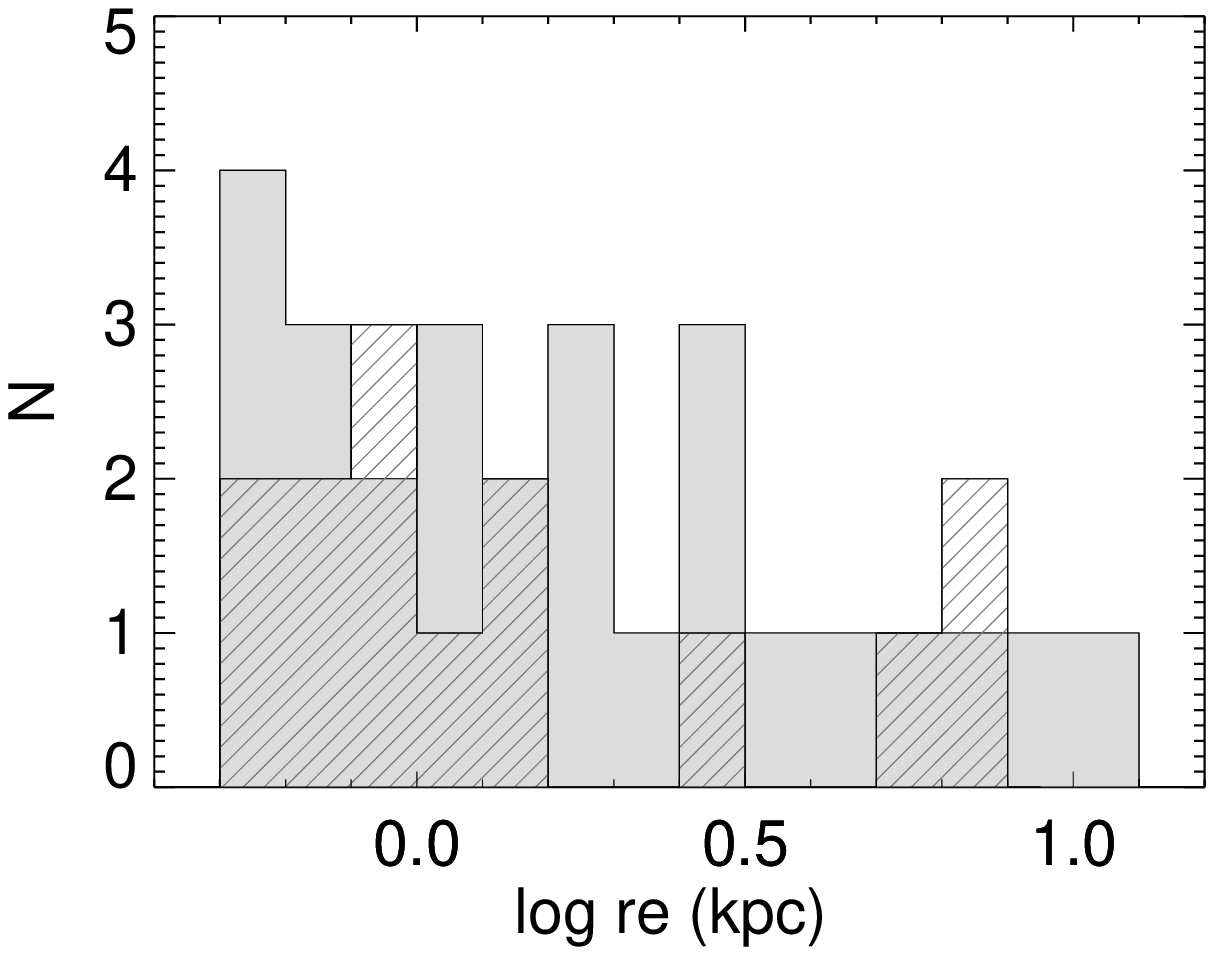}{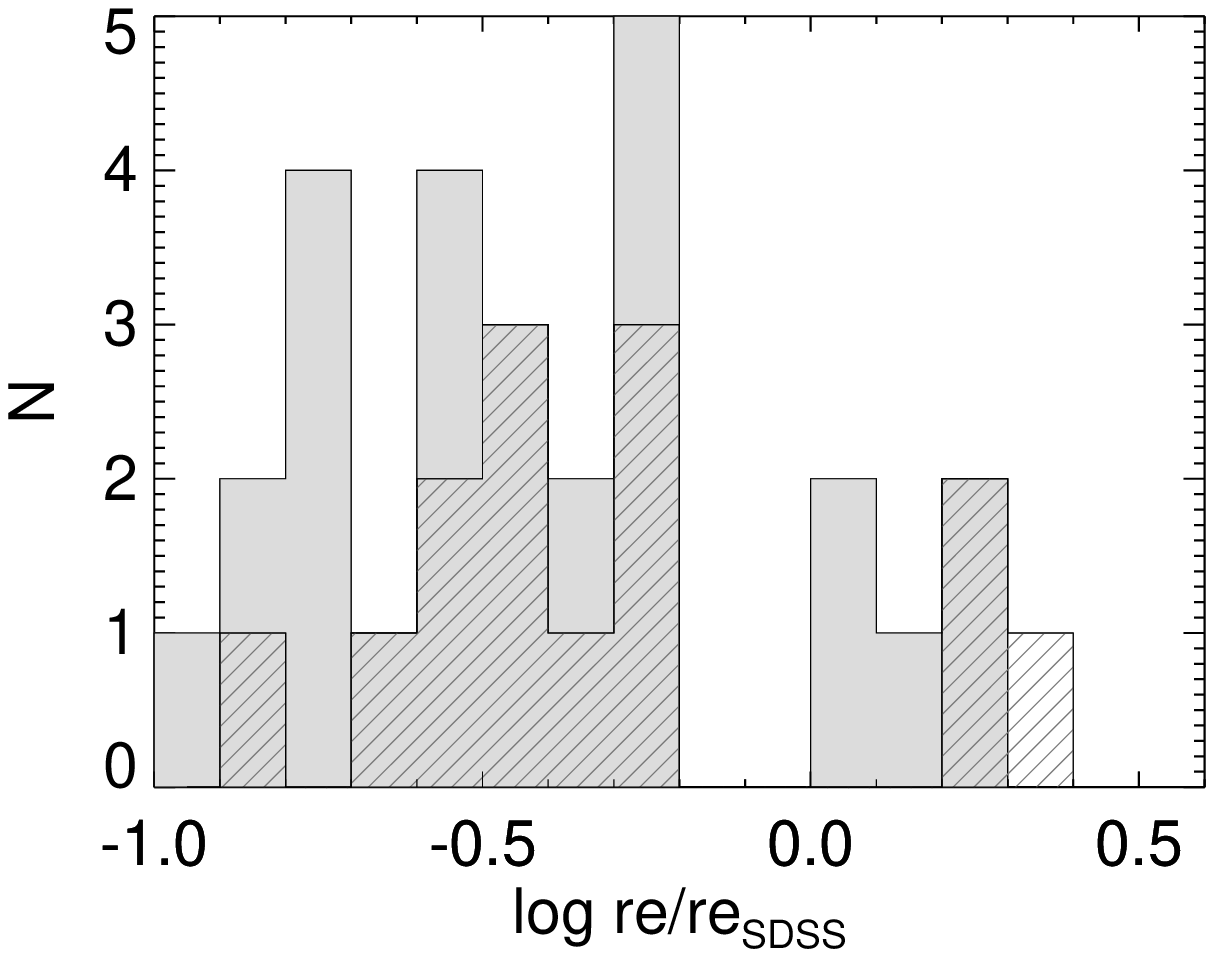}
\caption{Left: Distribution of derived circularized effective radii for DRGs (shaded) and DBGs (hatched). Right: Same as left plot, but normalized by the local mass-size relationship for late-type galaxies in SDSS \citep{shen2003}:  Each size has been divided by the size of a late type galaxy of similar mass in SDSS. In both plots, the distributions are not significantly different.  
\label{sizes_drgdbg}
\vspace{0.3cm}
}

\end{figure}

The $r_e$ distribution of the DBGs is similar to that of the DRGs. The KS-test yields a $64\%$ chance that the $r_e$ values were drawn from the same parent distribution.
In Fig.\ref{sizes_drgdbg} we take into account that the DBGs are on average not as massive as the DRGs by normalizing the sizes by the local mass-size relationship of local galaxies in the SDSS \citep{shen2003}. Note that we throughout the paper take into account the difference between the Salpeter IMF assumed here and the Kroupa IMF assumed by \cite{shen2003} by dividing our derived masses by two before comparing to \cite{shen2003}. As a possible caveat we note that \cite{shen2003} derive masses in a different than done here.
The mass-normalized comparison further supports the similarity of the size distributions. The KS-test yields a $50\%$ chance that the mass-normalized $r_e$ values were drawn from the same parent distribution. Furthermore it can be seen that the sizes of the galaxies in the sample are on average smaller by a factor of about three than late type galaxies of similar mass in the local universe (see Sec.\ref{masssize} for a more details).
Here, and in the following section we normalize all sizes using the local mass-size relation of late-type galaxies in \cite{shen2003}, as the vast majority of the galaxies in the sample are best fitted by exponential disk surface brightness profiles. Alternatively we could have normalized the sizes of galaxies with de Vaucouleurs like surface brightness profiles, by the local relation for early-type galaxies, however as the local early and late type relations are very similar in the mass range of the  ``de Vaucouleurs'' law galaxies in the sample (see Fig.\ref{massre}), it would not significantly change the normalized sizes. 

\begin{figure}
\epsscale{1.15}
\plottwo{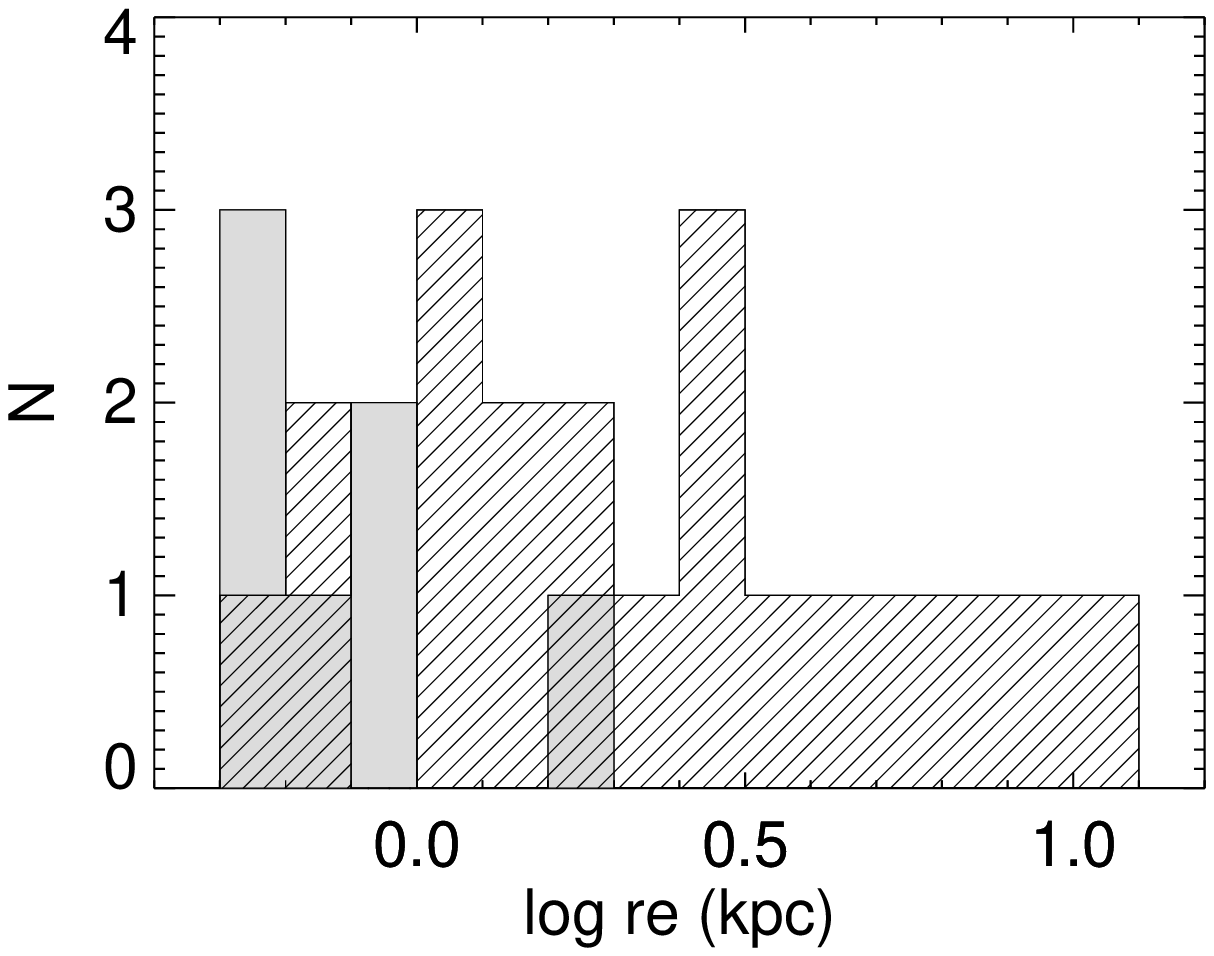}{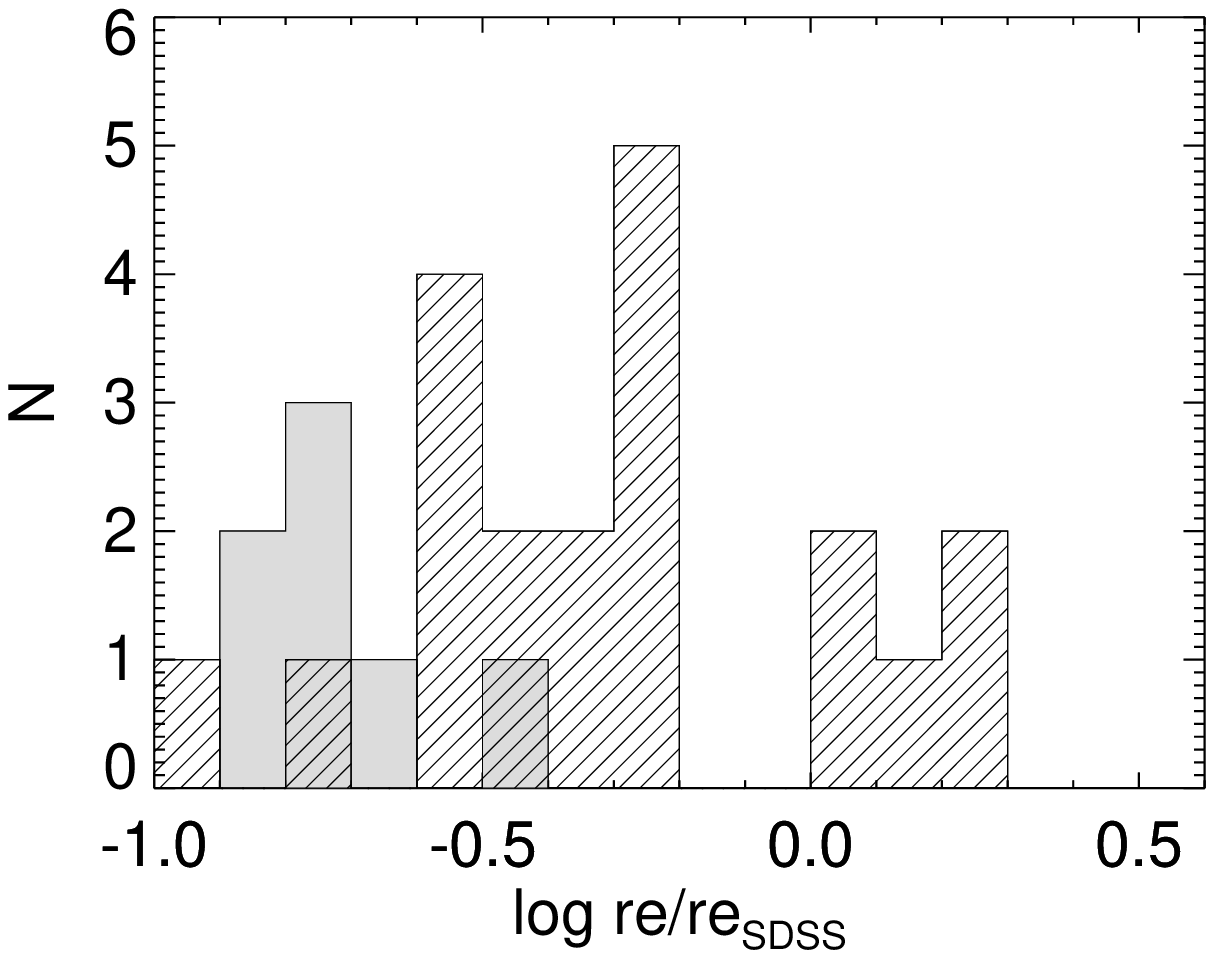}
\caption{Left: Distribution of derived circularized effective radii for quiescent (shaded) and star forming DRGs (hatched) histograms. Right: Same as left plot, but  normalized by the local mass-size relationship for late-type galaxies in SDSS \citep{shen2003}. The quiescent galaxies are significantly smaller than the star forming galaxies. 
\label{sizes_aq}
}
\end{figure}

While the sizes of the galaxies do not seem to depend on their J-K color, there is a strong correlation with star formation activity. In Fig. \ref{sizes_aq} we compare the size distribution of star forming and quiescent galaxies.  There is a clear trend that quiescent galaxies are smaller than star forming galaxies. Four of the quiescent DRGs ($57\%$) are unresolved in the NICMOS images, while none of the star forming DRGs are unresolved. As shown in Fig.\ref{sizes_aq} (right) the size difference is even more pronounced when normalized by the local mass-size relation.
From this figure it can be seen that the quiescent galaxies are on average smaller by a factor of about 5 than late type galaxies in the local universe, while star forming galaxies on average are smaller by a factor of about 2. 
The mass distribution of the star forming and quiescent DRGs are similar (the KS test yields a $80\%$ chance of them being drawn from the same parent mass distribution)
It will be interresting to determine whether star forming galaxies have lower mass densities or are dense systems with widely distributed star formation.
In Fig.\ref{n-distributions} we compare the n-distributions of the resolved star forming and quiescent galaxies. All of the resolved quiescent galaxies and $83\%$ of the resolved star forming galaxies are better fitted by exponential disk than de Vaucouleurs like surface brightness profiles. The KS test yields a $54\%$ chance of the n-values of quiescent and star forming galaxies being drawn from the same parent distribution.

\section{CORRELATIONS BETWEEN SIZE AND MIPS {24\micron} EMISSION}
\label{mipsvssize}
At $1.5<z<2.5$ the $24\micron$ MIPS filter samples rest-frame $\sim 6-10\micron$, and the $24\micron$ flux offers a powerful method for differentiating between galaxies with SEDs dominated  by dusty star formation and/or AGN, both of which produce substantial MIR emission, and quiescent galaxies whose stellar flux continues to drop longwards of $\sim 2 \micron$ \citep[e.g.][]{webb2006}. In pure starburst galaxies the MIR emission is dominated by PAH features, that are strong relative to the underlying dust continuum \citep[e.g.][]{smith2007}. The hard radiation field of an AGN destroys PAH carriers, and the continuum emission from hot small dust grains is strong throughout the MIR \citep{genzel2000, laurant2000}. 
At $z>2.5$ the sensitivity to star formation is reduced drastically as the PAH features are redshifted out of the 24$\micron$ band, so in the following we will concentrate on the galaxies in our sample with $z<2.5$, which leaves 19 DRGs and 4 DBGs. 
Assuming that the 24$\micron$ flux is dominated by star formation, the PAH and MIR emission can be used to estimate the current star formation rate of the galaxies \citep{wu2005}. Using an Arp 220 SED template, we extrapolated from $24\micron/(1+z)$ flux  to rest-frame $6.75\micron$ flux.   We then estimated the total ($8-1000\micron$) infrared luminosity $L_{IR}$ through the observed $6.75\micron-F_{FIR}$ relation calibrated locally with the Infrared Space Observatory \citep{elbaz2002}, which shows a factor of 2 scatter.  This value can then be converted into a SFR using the $L_{IR}-SFR$ relationship of \cite{bell2003}, with an expected scatter of at least a factor of two. Additional systematic uncertainties are introduced in the first step extrapolation. Adopting a M-82 like template instead of the Arp 220 template leads to a factor of two lower SFR \citep{webb2006}, while adopting the model templates of \citet{dale2002} can result in factors of 2-6 lower SFRs \citep{papovich2007}. Uncertainties in the photometric redshifts translates into a factor of 2.5 uncertainty in the derived SFR \citep{papovich2006}. Added  in quadrature the expected uncertainties in 24$\micron$-SFR conversion amount to about a factor of $\approx 7$. Furthermore, the SFRs derived in this ways are upper limits, as AGN may contribute to, or in some cases dominate the $24\micron$ flux, leading to overestimated SFRs.
\begin{figure*}
\epsscale{1.15}
\plottwo{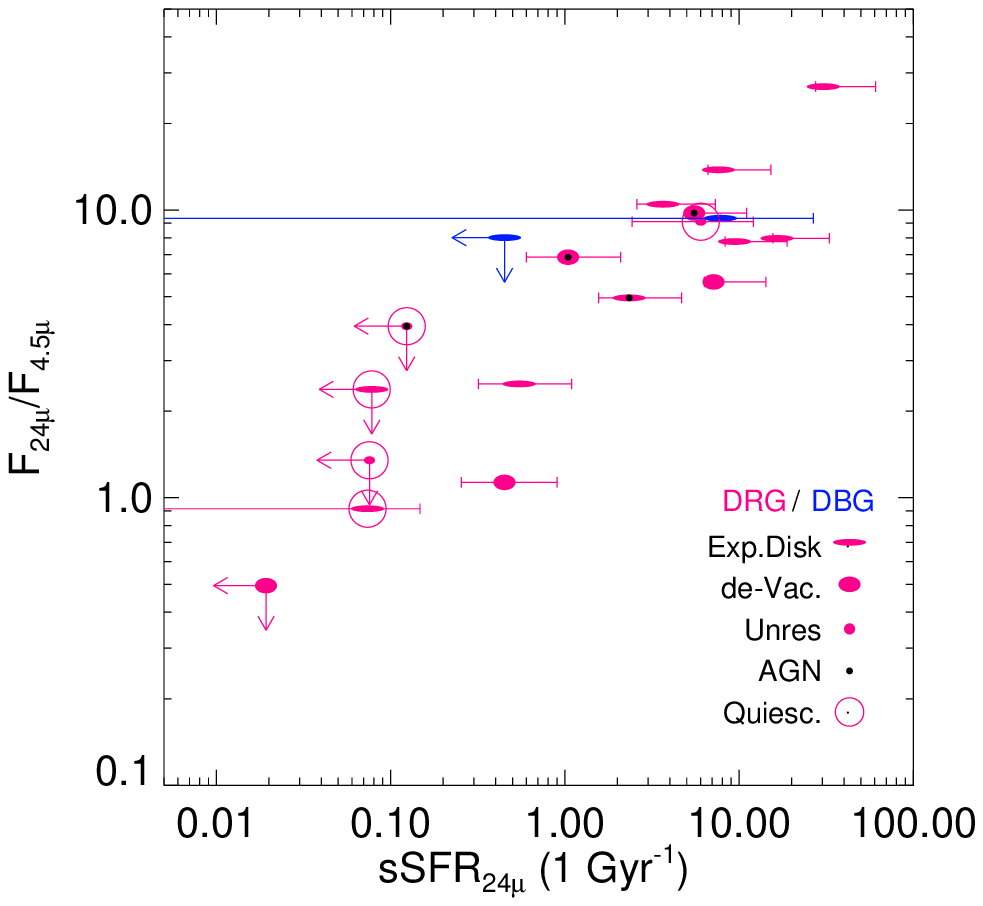}{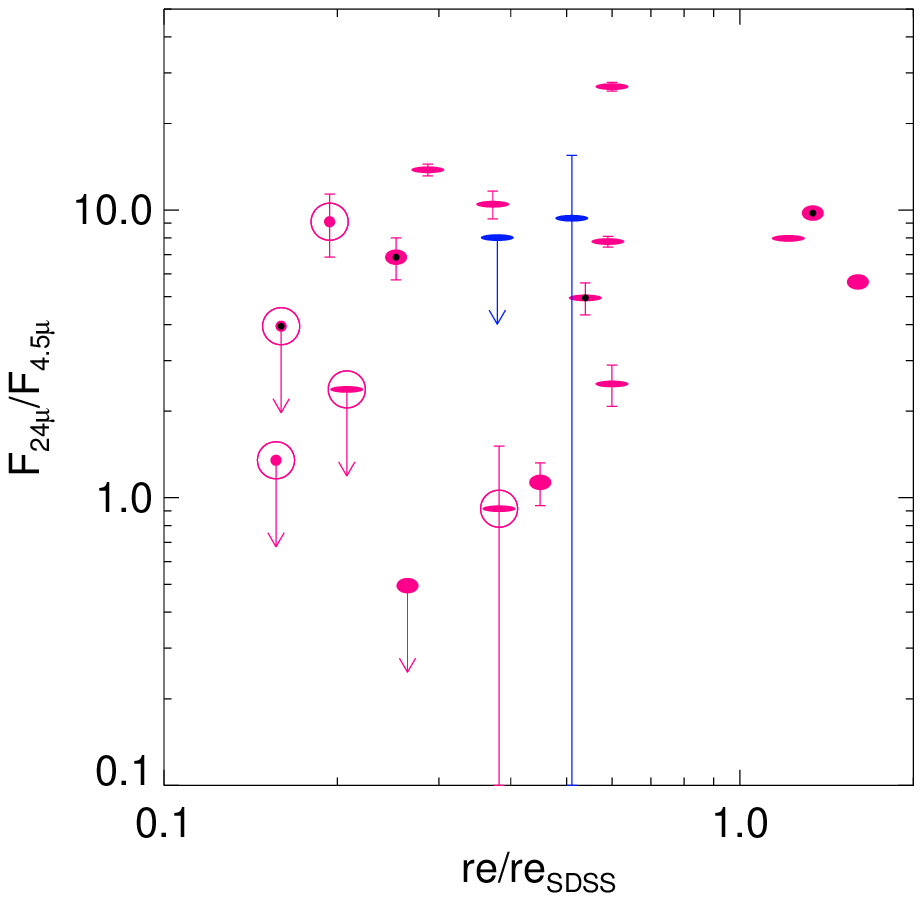}
\caption{Ratio of flux in the MIPS $24\micron$ and IRAC $4.5\micron$ bands, versus the sSFR derived from the $24\micron$ flux (left) and normalized size (right) for galaxies in the sample with $2\lesssim z_{phot}<2.5$. Error bars are one sigma and include only photon and confusion noise associated with the $24\micron$ photometry, not systematic errors related to the conversion from flux to SFR. For clarity errors on the sSFR are only shown in the left plot, and errors on $F_{24\micron}/F_{4.5\micron}$ are only shown in the right plot.  Galaxies with negative measured fluxes are shown at 1$\sigma$ upper limits (indicated by arrows).  There is a relation between the derived specific star formation rate and the $F_{24\micron}/F_{4.5\micron}$ ratio, galaxies with larger ratios have larger sSFRs. There is also a relation between $F_{24\micron}/F_{4.5\micron}$ ratio and size, galaxies with larger ratios on average have larger sizes.  
\label{ratios}
}
\end{figure*}
In Fig.\ref{ratios} (left) we plot the specific star formation rate derived from 24$\micron$ flux versus the ratio of flux in the MIPS 24$\micron$ and IRAC $4.5\micron$ bands. There is a clear correlation, galaxies with higher $F_{24\micron}/F_{4.5\micron}$ ratios have higher derived sSFRs.  This ratio is a purely observational quantity which does not depend on extrapolations and (only weakly, through the assumed photo-z) on assumptions about the SED, and thus is a good simple estimator of star formation/AGN activity in the galaxies. The origin of the relation is the fact that $F_{24\micron}$ roughly scales with the SFR (and redshift) of the galaxies and $F_{4.5\micron}$ (which samples the restframe J-band at $z\sim2.5$) correlates linearly with their stellar mass (with a considerable scatter).

Four of the five galaxies in the relevant redshift range (with IRAC coverage) which were classified as quiescent based on their optical-IRAC SEDs have very low inferred $sSFR_{24\micron}$ (consistent within 1$\sigma$ with no ongoing star formation) and  $F_{24\micron}/F_{4.5\micron}<1$ (within $1\sigma$), while one (DRG 5) formally has a higher derived sSFR and  $F_{24\micron}/F_{4.5\micron}$, but is only a $\sim 2 \sigma$ detection at $24\micron$.  Twelve of the thirteen galaxies (with IRAC coverage) which were classified as star forming have high $sSFR_{24\micron} >  0.3 Gyr^{-1}$ and $F_{24\micron}/F_{4.5\micron}>1$,  while one (DRG 25) is undetected at $24\micron$. Note that this galaxy which is almost equally well fit by the star forming and quiescent SED model (see Sec.\ref{sedmodel}.) 
Four galaxies in the relevant redshift range did not have IRAC coverage and are therefore not plotted in Fig.\ref{ratios}, but we note  that the two quiescent galaxies (DRG 20, DBG 3) has $sSFR_{24\micron}$ consistent within 1$\sigma$ with no ongoing star formation, while the two star forming galaxies (DRG 19, DBG 13) have  $sSFR_{24\micron}= 2.4, 11.4$ respectively.
We did not find evidence for a dependency of the derived $sSFR_{24\micron}$ on stellar mass in our sample.
We note that the $sSFRs$ for these galaxies are much higher than those of galaxies of similar mass at lower redshift. This suggests that a significant fraction of massive galaxies was still forming stars at $z\sim2.5$ whereas they are quiescent today \citep[see also ][]{vandokkum2004, papovich2006}.
In Fig.\ref{ratios} (right) we plot the $F_{24\micron}/F_{4.5\micron}$ ratio as a function of normalized size. Galaxies with low  $F_{24\micron}/F_{4.5\micron}$ ratios are significantly smaller than galaxies with large $F_{24\micron}/F_{4.5\micron}$ ratios. This provides independent evidence for the relation between star formation activity (derived from SED fits) and size presented in Sec.\ref{sec.sizes}.
While it is hard to disentangle the contribution of star formation and AGN to the measured $24\micron$ flux, the MIPS observations independently confirm that the galaxies classified as quiescent do not host significant ongoing star formation or AGN activity, and the interpretation that they are small and red due to very compact old stellar populations, rather than dust enshrouded central AGN.   In Sec.\ref{agn} we explore the presence of AGN in the full sample.
Note that 12/13 DRGs at $2<z<2.5$ with MIPS 24$\micron$ detections have inferred total infrared luminosities comparable to local Ultra Luminous Infrared Galaxies (ULIRGS, $L_{IR} \gtrsim 10^{12} L_{\odot}$) which place them among the most powerful starbursts known, in some cases with $SFR \gtrsim 1000 M_{\odot}$. This is consistent with the fact that two of the brightest DRGs at $24\micron$  (DRG 13 and 17) are also directly detected in JCMT+SCUBA $850\micron$ imaging \citep[see][]{knudsen2005}.
Similar high total infrared luminosities are derived by \cite{papovich2006} for DRGs in the same redshift range.    
   
\section{AGN}
\label{agn}
Previous observations of DRGs suggest that some of them host AGN.  Different techniques for detecting the presence of AGN find different fractions, mainly due to different selection biases, and sensitivities. Based on optical spectroscopy \cite{vandokkum2003b} find that 2 out of 6 DRGs with emission lines are consistent with AGN activity, corresponding to 33\%. Based on NIR spectroscopy, \cite{kriek_agn2006} find 4 out 20 K-selected galaxies between $2<z<2.7$  (20\%) with secure NIR spectroscopic redshifts host AGN. 
Based on relatively shallow X-ray observations (91 ks exposure with the Chandra X-ray telescope) of the MS1054-03 field, \cite{rubin2004} find that two out of 40 DRGs have x-ray point source counterparts, corresponding to a AGN fraction of $5\%$.  
Combining X-ray observations and the shape of the MIR continuum, \citet{papovich2006} found an AGN fraction of $25\%$ among a sample of 153 DRGs in the redshift range $1.5<z<3.0$. 

Here we estimate the AGN fraction in our sample and the influence on our results, by combining the X-ray imaging \citep{rubin2004}, with the shape of the MIR continuum, and visual inspection of the SEDs (for the presence of excess restframe UV flux) and structure (for the presence of blue central point sources).
Two of the DRGs in MS1054-03 field (DRGs 26 and 27) are detected as X-ray point sources, demonstrating that they host AGN, but in addition to these there may be other AGN which are too faint to be detected directly. None of the DBGs are detected in the rather shallow X-ray observations.

Another way to locate AGN is by inspecting the IRAC colors of the galaxies. AGN stand out in the $5.8-8.0$ vs $3.6-4.5 \micron$ color-color plot \citep[see fig.1 in ][]{stern2005}.
Adopting the empirical criteria of \cite{stern2005}, 5 of the DRGs in our sample have IRAC colors consistent with hosting AGN (DRGs 1, 6, 9, 21, 26), while two fall just outside the selection criteria (DRGs 10 and 27). 
Another indicator of AGN activity is when galaxies with $z>2.5$ (beyond which the sensitivity to starburst induced emission drops drastically) have significant MIPS detections. Three DRGs (8, 10, 26) fall in this category.

In Tab.\ref{modeltable} it is indicated which galaxies in the sample meets the different AGN selection criteria.
There is a large overlap between the subsamples selected from X-rays, IRAC and MIPS. Both X-ray selected candidates, and 2/3 MIPS selected candidates are included by the IRAC color selection criterion. 
The MIPS and X-ray data are only deep enough to detect the brightest AGN, so the three galaxies which are selected from IRAC, but not the other techniques could either host fainter AGN or not host AGN at all \citep[some contamination  ($<20\%$) is expected for the IRAC color selection method,][]{stern2005}. 

Two DBGs have IRAC colors consistent with hosting AGN (DBGs 10 and 12). DBG 10 is potentially an ongoing merger of two galaxies (with consistent photometric redshifts, see Sec.\ref{morphologies}), one of which has a blue point source in the center, which is likely an AGN. DBG 12 is a blue, very compact galaxy which is  unresolved in the NICMOS image, but resolved in the ACS image. Finally, one $z>2.5$ DBG is detected by MIPS (DBG 5). This galaxy is very blue in the optical, suggesting it could host an AGN.  

The estimated AGN fraction in the DRG sample is thus between $7-30\%$, depending on which selection method is used, and 15$\%$ in the DBG sample.

It is not straightforward to estimate the relative contribution of the possible AGN to the SED, but 
the most important point for the present analysis, is that 7 of the 8 galaxies which are candidates for hosting AGN are well resolved in the rest-frame optical, (with sizes spanning the full range 1-8 kpc), and are classified as star forming, based on their SEDs, indicating that the possible AGN does not dominate the rest-frame optical emission and structure. Inspection of the SEDs of the AGN candidates confirms that the rest-frame optical emission is well fitted by stellar population models (with significant spectral breaks), but in all cases there is some evidence for excess rest-frame UV emission with respect to the models, which could be attributed to AGN. The UV excess is however only significant for DRG 26 and 27 (the X-ray sources).  
Of the 8 AGN candidates, only one (DRG 1) has a central point-like sources in the stacked ACS image, suggesting that the possible AGNs in the remaining 7 galaxies are obscured by dust.  
Visual inspection of the SED of DRG 21, (the only AGN candidate that is unresolved in the NICMOS images), shows that it is well fitted by the SSP model with a strong $4000${\AA} break, but with excess flux in $8\micron$ band. Note that this galaxy is neither detected in X-rays or at $24\micron$. 
While the derived star formation rates, ages and masses, dust content etc., of the galaxies in question could be affected by the possible presence of AGN, it is not expected to significantly affect the conclusions of this study, as the main correlation between size and star formation activity persists even when the AGN candidates are excluded from the sample.     
 Furthermore, since the AGNs do not dominate the restframe optical emission of the galaxies, their influence on the derived SED fit parameters is likely modest.

\section{THE RELATION BETWEEN STRUCTURE AND STELLAR POPULATIONS}
\label{correlations}
In this section we explore the data for correlations between structural properties of the galaxies, and properties of their stellar populations derived from the SED fits.
\subsection{Mass - Size Relation}
\label{masssize}
\begin{figure*}
\epsscale{1.15}
\plottwo{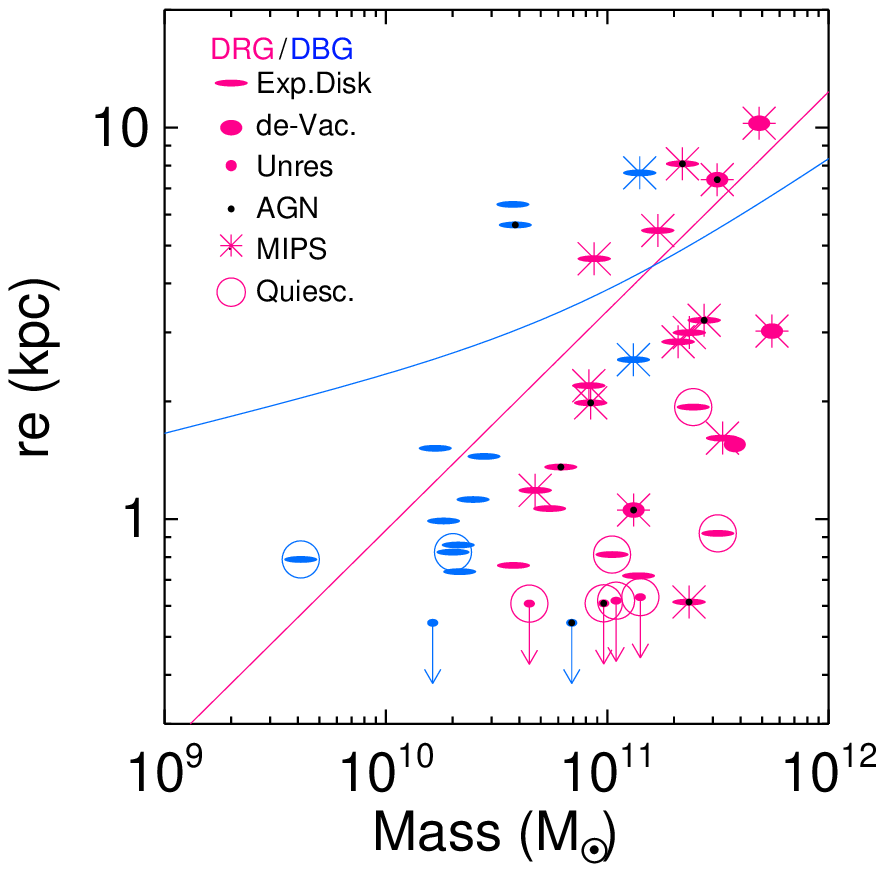}{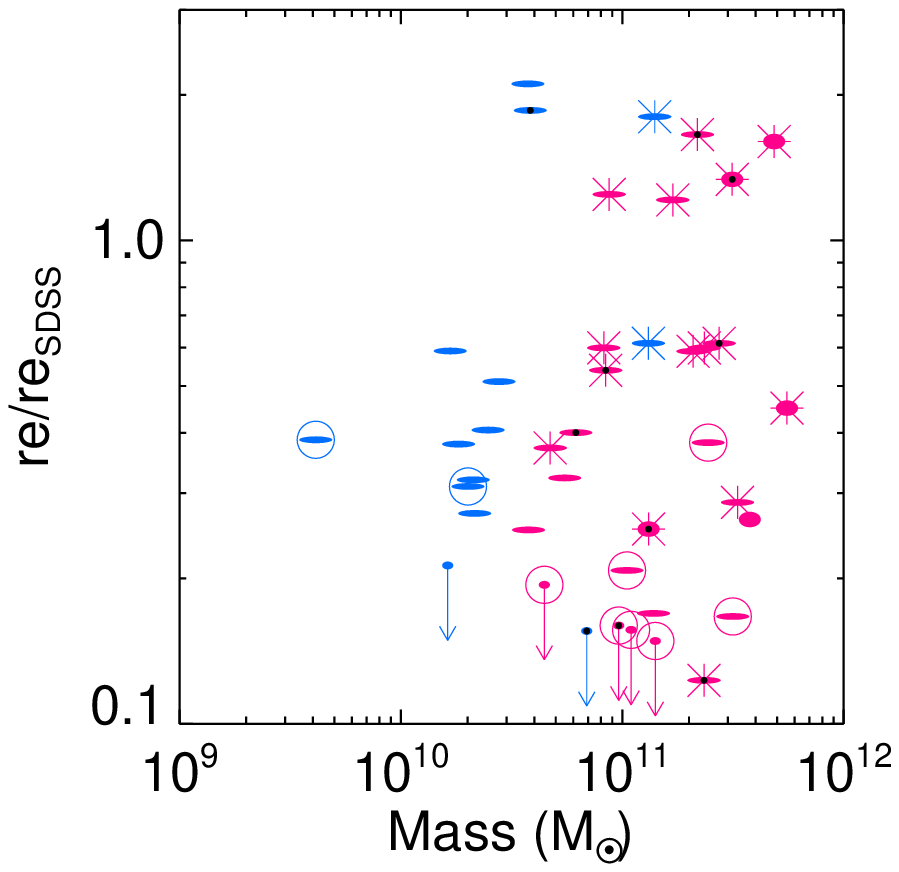}
\caption{Left: Sizes (effective radii) as a function of stellar mass. Red symbols are DRGs, blue symbols are DBGs. ``Flat'' symbols  have ``exponential disk like'' like surface brightness profiles, ``Elliptical'' symbols have  ``de Vaucouleurs-like'' surface brightness profiles. Small symbols are unresolved in the NICMOS images, and are represented by an upper limit on their size. Open circles indicate galaxies with SEDs consistent with being quiescent, the rest have SEDs consistent with being star forming. Stars indicate ($3\sigma$) MIPS $24\mu m$ detections. Black dots indicate that the galaxy may host an obscured AGN (see Sec.\ref{agn}).
The curves are the mass-size relations of early-type (red) and late-type (blue) galaxies in SDSS \citep{shen2003}.
Right: Sizes, divided by the mass-size relation for late type galaxies in the SDSS (the blue curve in the left plot).
Quiescent galaxies are on average a factor of 5 smaller than late type galaxies of similar mass in the SDSS, while star forming galaxies are smaller by a factor of 2.
}
\label{massre}
\end{figure*}

In Fig.\ref{massre} (left) the size-mass relation of the individual subgroups of galaxies are compared to each other and to the relations in the local universe (SDSS).
It is evident that most of the galaxies are smaller than galaxies of similar mass in the local universe. It is not possible to determine whether the mass-size relation of the galaxies resembles the local distribution of late type or early type galaxies better, but since the majority of the galaxies are better fitted by exponential disk profiles, we will compare to the mass-size relationship of late-type galaxies in the following (unless noted otherwise).  In Fig.\ref{massre} (right) the mass-size relation is normalized by the local mass-size relation of late type galaxies.

The median sizes of both DRGs and DBGs are smaller by about a factor of 2.5 compared to late type galaxies of similar mass in the local universe ($\left< re/re_{SDSS}\right >_{DRG}=0.37 \pm  0.08$, $\left< re/re_{SDSS}\right >_{DBG}=0.41 \pm  0.21$) 

A larger difference is found when comparing the median sizes of quiescent and star forming galaxies: the median size of the quiescent galaxies is a factor of $\sim 5$ smaller ($\left < re/re_{SDSS}\right >=0.19\pm0.03$) than late type galaxies of similar mass in the local universe, while the median size of star forming galaxies are only smaller by a factor of $\sim 2$ ($\left < re/re_{SDSS}\right >=0.45\pm0.15$).
Note that if we normalize the sizes of the quiescent galaxies using the local mass-size relation of early-type galaxies, the median size ($\left < re/re_{SDSS}\right >=0.23\pm0.04$ is still consistent with being a factor of 5 smaller than local early type galaxies of similar mass. 
The star forming galaxies actually divide into two distinct subgroups: the majority have ($\left < re/re_{SDSS}\right >=0.52\pm0.14$), but 8 galaxies have  $\left< re/re_{SDSS}\right >=1\rm{-}3$, and are thus larger than late type galaxies of similar mass in the local universe. These 8 galaxies all have disturbed structure in the NICMOS images (see Fig.\ref{drg}: DRG- 13 17, 19, 26, 27 and Fig.\ref{dbg}: DBG-5, 10, 11), and all are detected by MIPS.  

The observed evolution of the mass-size relationship is consistent with the evolution of the mass-size relation derived from the FIRES survey for galaxies at $z\sim2.$5 which were found to be a factor of $2\pm0.5$ smaller than galaxies of similar mass in the SDSS \citep[based on sizes derived from ground based data, ][]{trujillo2006}.     
Interestingly, from a study of the mass-size relation in the HDFN \citep{papovich2005} find that galaxies at $\sim2.3$  are ``only'' $40\%$ smaller than galaxies at $z\sim1$, so the majority of the size evolution must take place between $0<z<1$.

\subsection{Surface Mass Density}
\begin{figure*}
\epsscale{1.15}
\plottwo{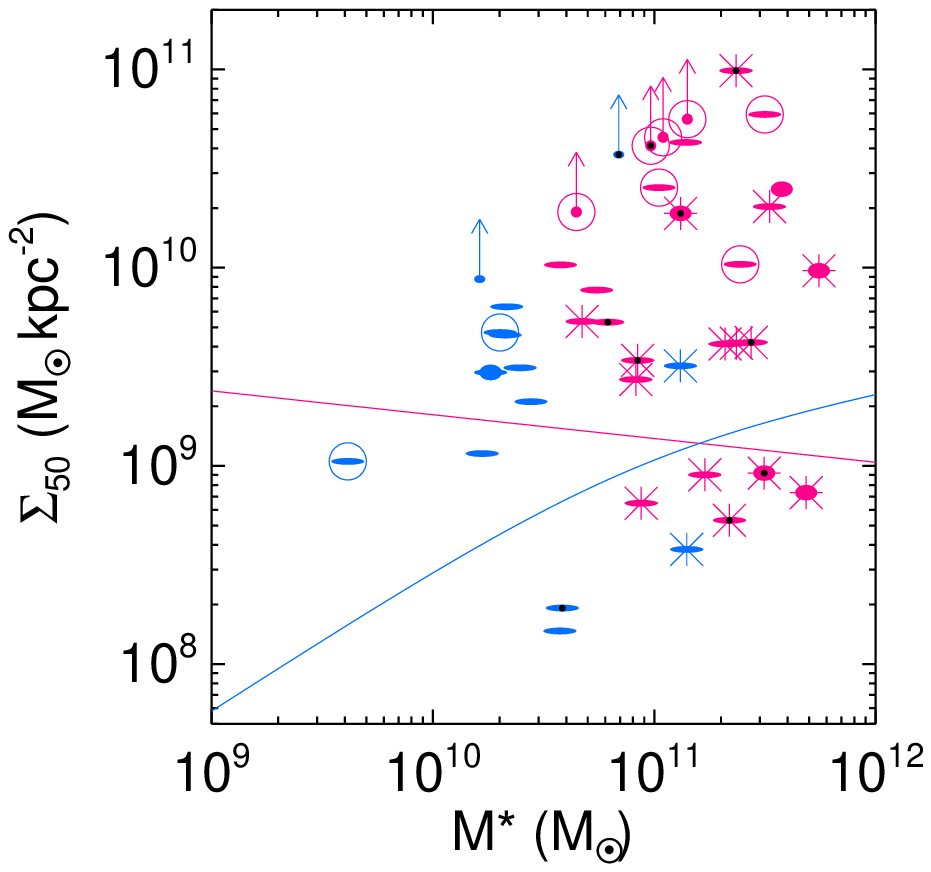}{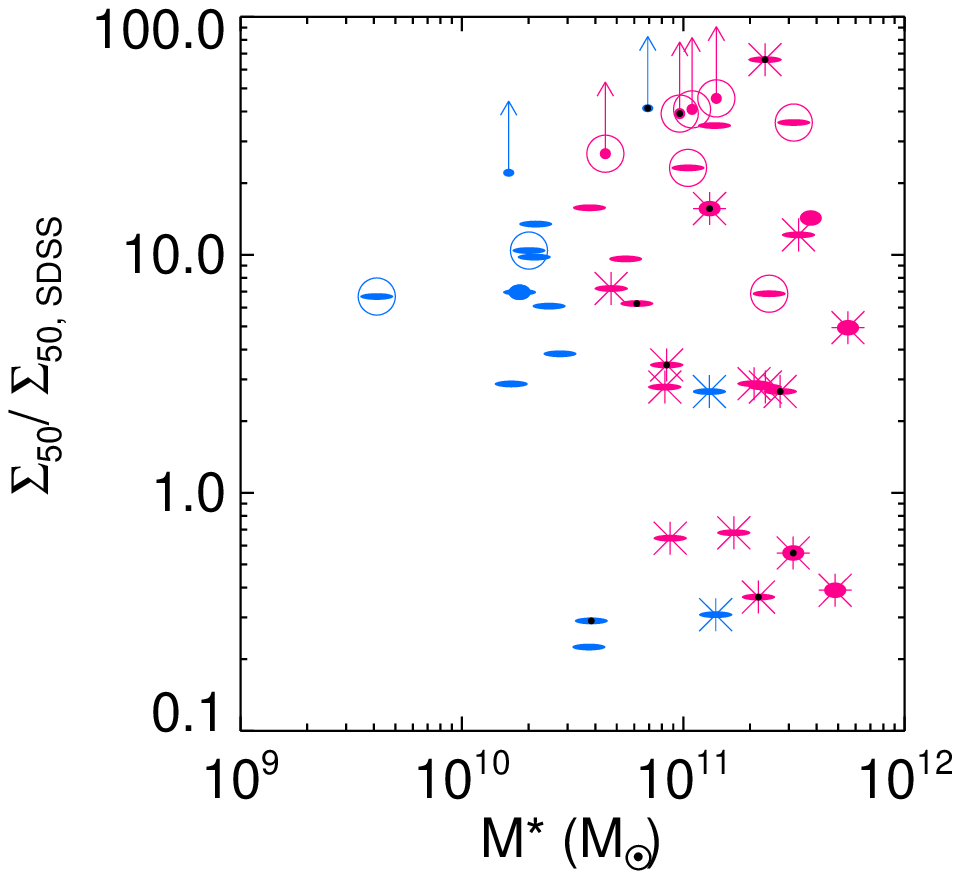}
\caption{Left: Average surface mass density within $r_e$ as a function of stellar mass. The lines are the local relations for late (blue) and early (red) type galaxies in SDSS calculated from the mass-size relations of \cite{shen2003}. Right: Surface mass density, normalized by the surface mass density of late-type galaxies in SDSS (the blue line in the left plot). The star forming/quiescent galaxies are on average 6/36 times denser than local late type galaxies of similar mass. Symbols are as in Fig.\ref{massre}    
\label{massdens}
}
\end{figure*}

In this section we combine the derived stellar masses and sizes of the galaxies in the sample, to derive surface mass densities. If we assume that mass follows light, the surface mass density of a galaxy within the effective radius is 
\begin{equation}
\Sigma_{50}=\frac{M_*/2}{\pi r_e^2}. 
\end{equation}
In  Fig.\ref{massdens} (left) we plot the derived surface mass density versus the total stellar mass, and compare to the mass densities derived for galaxies in the local universe in a similar way from the mass-size relations of \cite{shen2003}.
The galaxies follow a similar relation as for late-type galaxies in the local universe (with a large scatter), with more massive galaxies being denser, but shifted to higher mass densities for a given stellar mass. 
In Fig.\ref{massdens} (right) we normalize the mass densities by the local late-type relation. 
 In this plot it can be seen that there is no evidence for DRGs and DBGs following different M-$\Sigma$ relations. The DRGs are more massive than the DBGs, but have similar sizes, which translates into higher surface mass densities. The median surface mass densities are $\left < \Sigma_{50}/\Sigma_{50,SDSS} \right >_{DRG} = 9.6 \pm 4.2$ and  $\left < \Sigma_{50}/\Sigma_{50,SDSS} \right >_{DBG} = 6.1 \pm 2.4$ times higher than for late type galaxies of similar mass in SDSS for DRGs and DBGs respectively. 
The quiescent galaxies separate out from the star forming galaxies in this plot with much higher surface mass densities. The median surface mass density of quiescent galaxies is 1-2 orders of magnitude higher than in late type-galaxies in the local universe, with a median value of $\left < \Sigma_{50}/\Sigma_{50,SDSS} \right >_q = 36.0 \pm 5.8$ while the median surface mass density of the star forming galaxies is $\left < \Sigma_{50}/\Sigma_{50, SDSS} \right >_a = 6.2 \pm 3.2$. The star forming galaxies separate into two groups, the majority has surface mass densities around the median value, but 8 galaxies have surface mass densities which are lower than in late-type galaxies of similar mass in the local universe. These are the same galaxies which stood out in the mass-size plot (see Sec.\ref{masssize}). Note that if we normalize the sizes of the quiescent galaxies by the mass-size relation of local early type galaxies, the derived median  $\left < \Sigma_{50}/\Sigma_{50,SDSS} \right >_q = 33.4 \pm 9.9$ is still much higher than in the local early type galaxies of similar mass.

\subsection{Velocity Dispersions}
Here we assume that the galaxies have the same dynamical structure as local early type galaxies and are part of the same homologous family as local galaxies, hence their velocity dispersion can be estimated from their size and mass: 
\begin{equation}
 log(M_*)=2log(\sigma_{V})+ log(r_e)+6.07,
\label{fp}
\end{equation}
\citep{jorgensen1996, vandokkum03}.
The assumption of homology has been tested only out to $z\sim0.5$  \citep{vandermarel2006}, but the work of \cite{vanderwel2005} suggest that these relations hold at least out to $z\sim1.2$. However, as we shall see later it may not hold for the galaxies in our sample, as they will likely undergo significant structural evolution from $z\approx 2.5$ to the present \citep[see also][]{almeida2007}.

\begin{figure*}
\epsscale{1.15}
\plottwo{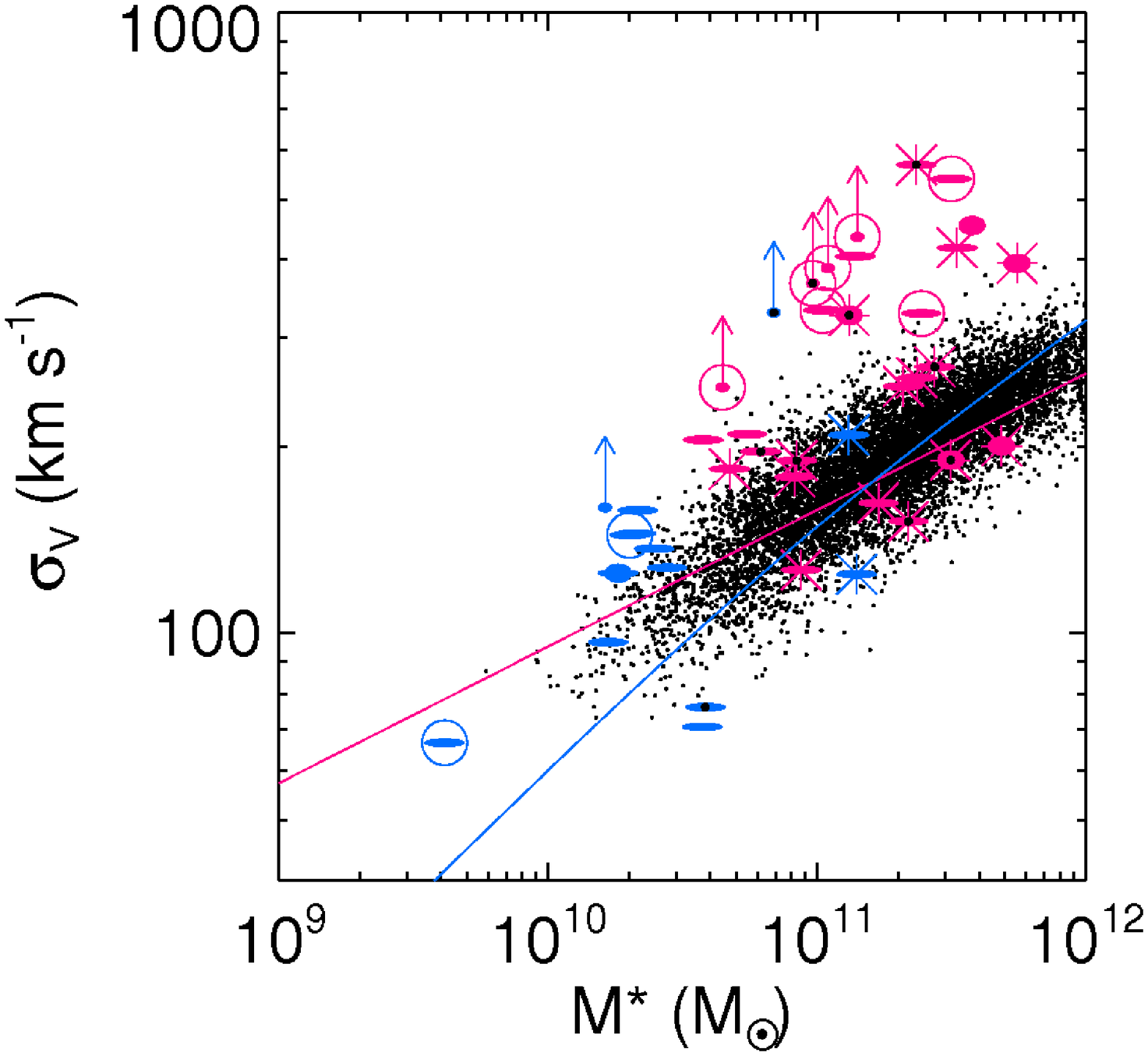}{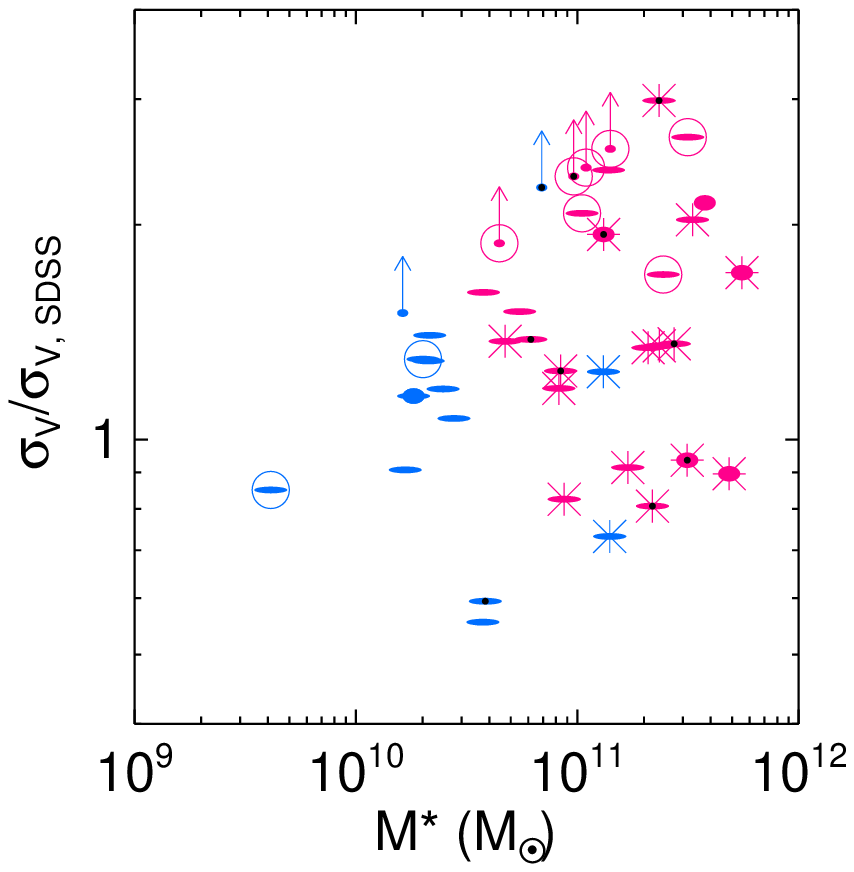}
\caption{Left: Velocity dispersion (estimated from the size and mass) of the galaxies versus their stellar mass. Small black dots are measured velocity dispersions of early-type galaxies in SDSS \citep{bernardi2003}, the curves are velocity dispersions calculated from the mass-size relations of \cite{shen2003} for late type (blue) and early type (red) galaxies in SDSS. The calculated relations agree well with the measured velocity dispersions from \cite{bernardi2003}.  Right: Calculated velocity dispersions divided by the velocity dispersion of late types in the SDSS (the blue line in the left plot). The star forming galaxies have velocity dispersions similar to galaxies of similar mass in the SDSS, but the quiescent galaxies have (0.5-3.5 times) higher derived velocity dispersions than galaxies of similar mass in the SDSS. Symbols are as in Fig.\ref{massre}
\label{disp}
}
\end{figure*}

\cite{bernardi2003} compiled a sample of early-type galaxies in the SDSS which were sufficiently bright to measure reliable velocity dispersions.  
In Fig.\ref{disp} we compare the velocity dispersions of local ellipticals in the \cite{bernardi2003} sample with those calculated for the galaxies in our sample. The stellar masses of the \cite{bernardi2003} galaxies were calculated from their $r_e$ and $\sigma_{V}$ using Eq.\ref{fp}. We also compare to the velocity dispersions of late-types and early types calculated from the mass-size relations of \citep{shen2003} using Eq. \ref{fp}. 
The quiescent and star forming galaxies fall in separate parts of the diagrams. The star forming galaxies roughly follow the local relation with a large scatter (but not much larger than the scatter of the local relation). The median velocity dispersion for a given stellar mass for the star forming galaxies is $\left< \sigma_{V}/\sigma_{V, SDSS}\right > = 1.4 \pm 0.1$ times higher than for galaxies of similar mass in the SDSS.
The quiescent DRGs have significantly (1.8-2.8 times) higher derived velocity dispersions for a given stellar mass than for local galaxies (median $\left< \sigma_{V}/\sigma_{V, SDSS}\right > = 2.4 \pm0.2  $). 
Interestingly the quiescent DBGs do not have significantly higher velocity dispersions than the star forming galaxies of similar mass.     
The highest of the inferred velocity dispersions of the quiescent DRGs $250\rm{-}550 km/s^{-1}$ are higher than for any readily identifiable object in the local universe. Note however that possible systematic errors of up to a factor of two in the derived masses, could cause the derived velocity dispersions to be overestimated by a similar factor (see Sec.\ref{caveats}).   Implications of the high derived velocity dispersions are discussed in Sec.\ref{models}.

\subsection{Kormendy Relationship}
\begin{figure}
\plotone{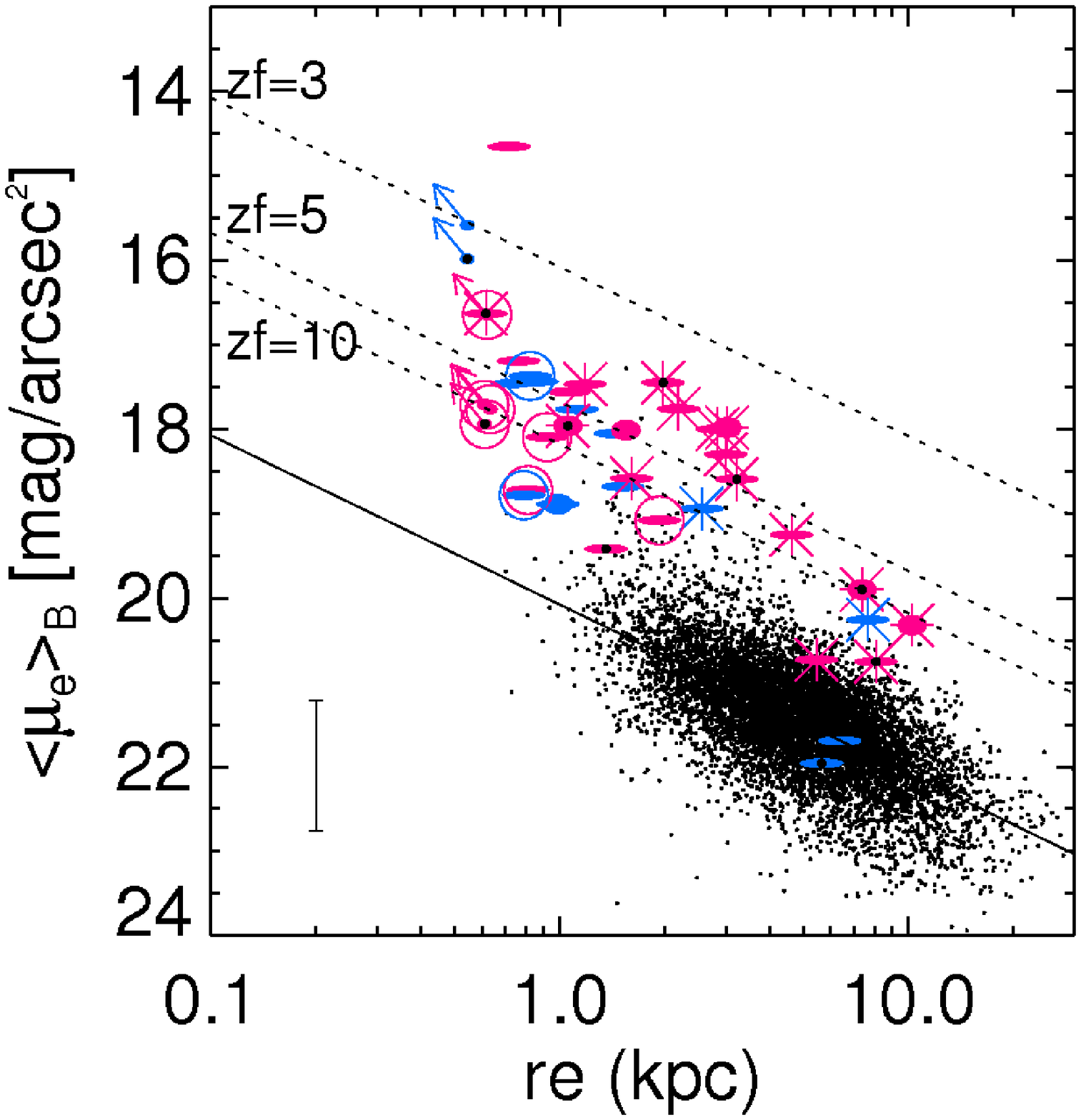}
\caption{Mean (dust extinction corrected) rest-frame B-band surface brightness within $r_e$, as a function of $r_e$ (the Kormendy relation).  The small points represents the Kormendy relation of early type galaxies in SDSS \citep{bernardi2003}. The solid line is the best fit relation to the SDSS points. The dotted lines are the expected relation for galaxies at $z=2.5$ for galaxies formed at $z_f=3,5$ and 10, which evolve passively (calculated from a single burst model of BC2003) from the formation redshift to $z=0$ (i.e no size evolution or change in slope of the relation). We note that the $z_f=5$ model fit the points well, but the size distributions of the $z\sim2.5$ and SDSS galaxies are quite different. Symbols are as in Fig.\ref{massre}.  For clarity, error bars (which are are dominated by uncertainties in the dust corrections) are not shown in the plot, but a typical size error bar is displayed in the lower left corner of the plot. 
\label{sb}
}
\end{figure}

In Fig.\ref{sb} we plot the surface brightness of the galaxies versus their size (the Kormendy relation). The rest-frame B-band dust corrected surface brightness of the galaxies has been calculated as:
\begin{equation}
\left < \mu_e \right >_B = b+5 log(\pi r_e)+ 5 log(2) -A_B - 2.5 log[(1+z_{ph})^4], 
\end{equation}
 where $b$ is the rest-frame B-band magnitude, 
derived from a linear interpolation between observed filters \citep[see][]{rudnick2003},  $r_e$ is the effective radius (in arcsec),
$A_B$ is the rest-frame B-band extinction (calculated using the \cite{Calzetti} extinction law assuming $R_V=4.05$ and the best fitting $A_V$ from the SED fit), and $z_{ph}$ is the photometric redshift.  

Also plotted is the Kormendy relation of early type galaxies in the SDSS \citep{bernardi2003}. The $B$-band surface brightness is estimated by extrapolating the observed $g$ and $r$-band surface brightnesses:  $\left < \mu_e \right >_B= \left < \mu_e \right >_r+1.97 (\left < \mu_e \right >_g-\left < \mu_e \right >_r)-0.25$ \citep{jorgensen1995}. The solid line is the best fitting relation, and the dotted lines are the expected relation at $z=2.5$ for galaxies formed at $z=3,5$ and 10 assuming that their surface brightness evolves passively from their formation redshift to $z=0$ (e.g assuming the slope of the local relation and  neglecting size evolution).
The Kormendy relation of the $z\sim 2.5$ galaxies shows a large scatter but roughly follows a relation with similar slope as the early-type galaxies in SDSS but shifted to brighter surface brightnesses and smaller sizes.
The zero point of the observed relation is consistent with the $z_f=5$ passive evolution prediction. We note however that the relatively large errors and scatter of the observed surface brightnesses makes it hard to put stronger constraints on the formation redshift, than $z_f \gtrsim 4$. 
Assuming passive luminosity evolution only, the Kormendy relation thus implies that the dominant stellar populations in these galaxies formed 1-2 Gyr earlier, when the Universe was $\lesssim 10\%$ of its present age.
Luminosity weighted formation redshifts of  $\sim 5$ are consistent with 
ages for quiescent DRGs derived from NIR spectroscopy \citep[0.5-1.5 Gyr, ][]{kriek2006}. 

This interpretation is however too simplistic as the sizes likely do evolve (as shown in Sec.\ref{masssize}). Many of the galaxies do not have any local counterparts to evolve into. Simple passive evolution would result in galaxies which are too small and too bright for their mass. 
Additional mass and size evolution (probably merger driven) is needed to explain the data, but the fact that the $z_f=5$ passive evolution model fits the Kormendy relation, may suggest that this evolution will be mainly be along the mass-size relation.

\section{COMPARISON TO MODELS}
\label{models}
We have compared the properties of the $z\sim2.5$ galaxies in the sample with properties of galaxies in the SDSS, and demonstrated that there is no simple passive evolutionary link between the two populations. To speculate on how and when these galaxies formed and what they will evolve into in the local Universe (and how?), we must turn to models. Current self-consistent models of galaxy formation and evolution have a hard time reproducing the properties of DRGs, so at the moment we can only do a qualitative comparison.    
DRGs dominate mass selected samples at $z\sim2.5$ \citep{vandokkum2006}. The most massive galaxies are believed to form in the highest density regions of the universe, which are the first to collapse, so DRGs are good candidates for being among the first galaxies to form. Also, since the DRGs at $z\sim2.5$ are already as massive as local ellipticals, they are likely to eventually end up as ellipticals at lower redshifts. But before they do, they probably have to go through a number evolutionary steps, including major and minor mergers. 
As for a possible formation scenario, semi-analytical modeling \citep[e.g.][]{khochfar2006} suggests that elliptical galaxies formed in gas-rich mergers can result in very dense stellar cores, as the gas is driven to the center of the merger where it produces massive starbursts. Galaxies which merge in the early universe are more likely to be gas-rich than galaxies which merge later, since they have had less time to exhaust their gas reservoirs, and as a consequence, massive ellipticals ($M>5\times 10^{11}M_{\sun}$) are predicted to be $\sim4$ times smaller at $z=2$ than galaxies of similar mass at $z=0$ \citep{khochfar2006}. The effect is smaller for less massive galaxies ($M\ga1\times 10^{11} M_{\sun}$) which are predicted to be $\sim 35\%$ smaller than galaxies of similar mass in local universe.  
High redshift gas-rich mergers are thus a likely formation scenario for the compact old quiescent $z\sim2.5$ DRGs. 
Possible examples of progenitors of quiescent DRGs are sub-millimeter galaxies (SMGs) which are believed to be dusty starbursts in ongoing gas-rich major mergers. CO observations have shown that these galaxies typically have dynamical masses $\sim 10^{11} M_{\sun}$ (mainly molecular gas and stars) within 2 kpc \citep{greve2005, tacconi2006, bouche2007}, corresponding to large surface mass densities (though not quite as high as for quiescent DRGs).    
How the quiescent galaxies relate to galaxies in the local universe is unclear, but they have to go through processes which increase their size, without increasing their mass too much, or creating too much new star formation, since their observed sizes are smaller and their surface mass densities and derived velocity dispersions are higher than for any known massive galaxy in the local universe. 
Dissipationless (``dry'') merging of gas poor systems could partly be responsible for this process \citep[e.g.][]{vandokkum2005, nipoti2003, boylan-kolchin2006}. Models suggest that such ``dry mergers'' could decrease the velocity dispersion and increase the size of the resulting remnant \citep{nipoti2003,  boylan-kolchin2006}. In order to reach the size and surface mass densities of local ellipticals the quiescent galaxies would however have to go through successive equal mass dry mergers, ending up with masses $\gtrsim 10^{12}M_{\sun}$. It seems unlikely that there would be that many progenitors of extremely massive galaxies in such a small field \citep[see discussion in ][]{zirm2007}.  It is more likely that they undergo both minor and major mergers, also with more gas-rich systems, which are more abundant. Whether such mergers are able to create an evolutionary link between the $z\sim2.5$ galaxies and local galaxies of more reasonable masses (i.e. $M< 10^{12}M_{\sun}$)  are yet to be explored by models.

The star forming DRGs may be easier to reconcile with local galaxies as they are ``only'' a factor of two smaller than galaxies with similar masses in the SDSS.
Models for the inside-out growth of disks galaxies through gravitational infall of gas \citep[e.g.][]{Bouwens2002, somerville2006} could account for a factor of two growth in size since $z\sim2.5$, but this is an unlikely scenario since the large masses of the DRGs may produce too many massive disk galaxies in the local universe. 
Additional processes like merging and fading of the young stars can also change the structure significantly.

\section{UNCERTAINTIES AND CAVEATS}
\label{caveats}
The main result in this paper is the correlation between size and star formation activity. To investigate the robustness of this result we here consider the uncertainties and possible caveats of the analysis. 
It is evident from the NICMOS images that the sizes of the quiescent galaxies are comparable to the size of the NICMOS PSF, so that sets a firm upper limit on their sizes. We have demonstrated that the systematic uncertainties in determining the sizes of the smallest galaxies are less than $15\%$ (mainly due to PSF variations). One could argue that the NICMOS observations may fail to detect larger very low surface brightness components of the galaxies, but this is unlikely, as no such components were found for DRGs in the much deeper Hubble Ultra Deep Field (HUDF) observations \citep{Toft2005}. Also, sizes derived from the independent very deep ground-based VLT+ISAAC FIRES data have been shown to agree well with the sizes derived here (see Sec.\ref{profiles}). 
Systematic redshift errors are perhaps the most important potential source of uncertainty, since these are important both for conversion from angular size to physical size, and for the properties of the stellar populations derived from the SED fits.
The faint optical magnitudes of DRGs make it very hard to obtain spectroscopic redshifts, so most of the work on DRGs (including the present) is based on photometric redshifts.  Comparison between spectroscopic and photometric redshifts suggests that the photometric redshifts are accurate to $\Delta z/(1+z) \simeq 0.08$ \citep{Forster-Schreiber2006} for galaxies  with emission lines in the optical, and recent work based on NIR spectroscopic observations of 4 DRGs suggests that the photometric redshifts are also accurate for DRGs without emission lines \citep{kriek_break2006}. The mean accuracy of the photometric redshifts of the four galaxies in the present sample with spectroscopic redshifts (see Tab \ref{sizetable}) is $\Delta z/(1+z) \simeq 0.09$. The important thing is that the number of catastrophic failures is small, so that low mass dusty $z\sim1$ galaxies or high redshift quasars are not mistakenly classified as compact $z\sim2.5$ massive quiescent galaxies. A large number of misclassified lower redshift galaxies is unlikely (and a few would not change the conclusions),  and the latter can be ruled out since the presence of high redshift QSOs would be evident from the IRAC and MIPS observations.
The uncertainties in the derived quantities (physical size, mass, surface mass-density etc) due to photometric redshift uncertainties were estimated in the following way: We changed the photometric redshifts to the extreme values allowed by the photometric redshift error bars and repeated the analysis, i.e. calculated their physical size, fitted their SED and derived their mass and other properties from the best fitting models. The median error due to known uncertainties in the photometric redshifts is in this way calculated to be $\sim 17\%$ for the physiscal sizes, $\sim 11\%$ for the masses, $\sim 15\%$ for the velocity dispersions and $\sim 40\%$ for the surface mass densities.
     
The stellar mass of the galaxies is an important parameter, both for the normalization of the size distributions using the local mass-size relationship, and for the interpretation of the $z\sim2.5$ $M-r_e$, $M-\Sigma$ and  $M-\sigma$ relations.  The most important sources of uncertainties in the mass estimates are the photo-z uncertainties discussed above, and uncertainties about the validity of the assumed Salpeter IMF. Assuming a Kroupa IMF rather than a Salpeter IMF results in masses which are smaller by a factor of two. Also, the present masses are derived from the best fitting BC2003 model. Using another stellar population synthesis model \citep[e.g.][]{maraston2005} could systematically shift the derived masses to lower values \citep[by about $\sim 40\%$, see][]{Wuyts2007}.

As the mass distribution of the quiescent and star forming galaxies are similar, however,  only systematic mass errors which are different for the two subgroups can change our conclusions, e.g if the masses of the quiescent galaxies were grossly over-estimated, and the masses of the star forming galaxies were under-estimated. A systematic shift of the whole sample, e.g. to smaller masses may make the evolutionary trends a bit less significant, e.g for the size evolution and derivatives thereof (surface mass-density and velocity dispersions), but would not change the conclusions.

Another possible source of systematic uncertainty is the possibility that selection effects prevent the detection of large, quiescent, low surface brightness galaxies. 
It is however unlikely that a significant number of such galaxies were missed, given that no such galaxies was found in the much deeper FIRES HDFS observations \citep{Labbe2003}, in the HDFN \citep{Dickinson2000} or in the HUDF \citep{Toft2005}. Furthermore, as shown by \citet{bouwens2004} in a study based on the UDF, the main effect of increasing the depth is to add galaxies at fainter magnitudes, not larger sizes, demonstrating that high redshift galaxies are predominantly compact, and that large low surface brightness objects are rare.
 
\section{SUMMARY}
\label{summary}
We presented HST NICMOS and ACS observations of 41 $2\lesssim z_{phot} \lesssim 3.5$ galaxies in the FIRES MS1054-03 field. Combining the FIRES optical-NIR data with deep Spitzer IRAC data we constructed optical-MIR SEDs,  which we fitted with synthetic stellar population models, to constrain the star formation histories, masses, ages and dust content of the galaxies.
Using NICMOS images, we performed a detailed structural analysis and showed that most of the DRGs are dominated by a single centrally symmetric component. About half of the DRGs are compact and half are extended. The DRGs are extremely faint in the optical, many of them are not detected, or detected at low S/N in the stacked 12h ACS exposure, and the ones that are detected frequently look very different than in the NICMOS images. We quantified this difference by calculating the shift of the centroid of the light between the ACS and NICMOS images. This quantity was shown to depend on J-K color, with the reddest galaxies having the largest shifts.
The DBGs on the other hand are bright in the optical, and have similar ACS and NICMOS structure.
The structural analysis suggests that the DRGs have composite stellar populations with old red stars dominating the stellar mass and rest-frame optical structure, in some cases with substantial amounts of ongoing star formation taking  place in off center regions of the galaxies. 
The DBGs on the other hand are consistent with simpler, uniformly younger stellar populations which dominate the structure at both rest-frame UV and optical wavelengths. 
The distribution of the Sersic parameters $n$ and $r_e$ of the DRGs and DBGs was shown to be similar. The galaxies in both subsamples have uniformly smaller sizes than galaxies of similar mass in the local universe (by a factor of about 2) and the majority of the galaxies ($\sim 80\%$) have surface brightness profiles which are better fitted by ``exponential disk'' like laws, than ``de-Vaucouleurs'' like laws.  These results are very similar to those derived from galaxies at $2<z<3.2$ from the FIRES survey \citep{trujillo2006}.   

Next we divided the sample into ``star forming'' and ``quiescent'' galaxies, based on stellar population synthesis model fits to their restframe UV-NIR SEDs.
The validity of this classification was confirmed by MIPS $24\micron$ observations, which showed that all except one of the quiescent galaxies had MIPS fluxes consistent with no ongoing star formation and/or AGN activity, while all except one of the star forming galaxies had MIPS fluxes consistent with significant star formation and/or AGN activity.   

The quiescent galaxies which constitute $22\%$ of the sample, have systematically smaller sizes than the star forming galaxies. The star forming galaxies follow a mass-size relationship (with a large scatter) similar to in the local universe, but shifted to smaller sizes by a factor of about 2 at a given mass. The median size of the quiescent galaxies is a factor of 5 smaller than galaxies of similar mass in the local universe, indicating a direct correlation between star formation activity and surface mass density at $z\sim2.5$.

The work presented here confirms, and adds considerable confidence, to the evidence for a relation between star formation activity and size which was found  in the HDFS \citep{zirm2007} through a much larger sample of galaxies (with a larger range of colors and masses), and the addition of MIPS $24\micron$ data to the analysis, which provides independent constraints on star formation and AGN activity.
  
The small sizes of the quiescent galaxies, combined with their high masses and luminosities implies that they have unusually high surface mass densities, surface brightnesses and, consequently high velocity dispersions.              
While the median surface mass density of the star forming galaxies is a factor of 6 greater than in galaxies of similar mass in the local universe, the surface mass densities of quiescent galaxies are more than an order of magnitude higher than in galaxies of similar mass in the local universe.
The velocity dispersions derived from the masses and sizes of the star forming galaxies are consistent with the velocity dispersions measured for galaxies of similar mass in the local Universe (taking into account a large scatter), but for the quiescent galaxies we derive velocity dispersions which are a median factor of  $>2$ higher. 
Finally we show that the Kormendy relation of the galaxies show a large scatter but follow a relation with similar slope as for early types in the local universe, and a zero point consistent with a mean stellar formation redshift of $z_f=5$. Combining the evidence from the structural and stellar population modeling analysis presented in this paper we conclude that in addition to passive aging of the stars, significant mass, size and structural evolution is needed for the $z\sim2.5$ galaxies to evolve into the local population.  

Our understanding of the compact quiescent galaxies will benefit greatly from future higher resolution NIR imaging studies (e.g. with WF3 on the HST) of larger samples, and NIR spectroscopic observations, providing spectroscopic redshifts, and stronger constraints on stellar populations.

\vspace{0.5cm}
Support from NASA grants HST-GO-10196.01-A and Spitzer 1268141 and from the Lorentz Center is gratefully acknowledged.

\input{tab1.tex}
\input{tab2.tex}

\end{document}

%% file: tab1.tex
\begin{deluxetable}{cccccrrc}
\tablecolumns{8}
\tablewidth{0pc}
\tablecaption{SED models }
\tablehead{
\colhead{Id} & \colhead{FIRES-Id} & \colhead{SED-type} &  \colhead{Mass} & \colhead{zph} &\colhead{$F_{24\micron}$} &\colhead{$F_{4.5\micron}$} & \colhead{AGN$^{(a)}$} \\
\colhead{} &\colhead{} &\colhead{} & \colhead{[$10^{10}M_{\sun}$]} & \colhead{} & \colhead{[$\mu$Jy]} & \colhead{[$\mu$Jy]} & \colhead{}
}
\startdata
     1 &    1644 & Star forming &     8.41 &      2.2 &     59.7$\pm$    13.4 &     12.1$\pm$     0.4& I \\
     2 &    1712 & Star forming &     4.72 &      2.0 &     69.5$\pm$    13.5 &      6.6$\pm$     0.4& \\
     3 &    1612 & Quiescent &    24.42 &      2.2 &     12.4$\pm$    14.2 &     13.5$\pm$     0.4& \\
     4 &    1734 & Quiescent &    31.53 &      2.7 &      6.9$\pm$    13.6 &     23.5$\pm$     0.2& \\
     5 &    1038 & Quiescent &     4.44 &      2.4 &     45.0$\pm$    19.6 &      4.9$\pm$     0.4& \\
     6 &    1050 & Star forming &     6.16 &      3.0 &     39.0$\pm$    19.2 &      3.3$\pm$     0.4& I \\
     7 &    1144 & Quiescent &    10.50 &      2.2 &      -31.8$\pm$    13.4 &      5.6$\pm$     0.4& \\
     8 &    1198 & Star forming &    23.38 &      2.8 &     95.8$\pm$    15.5 &      7.7$\pm$     0.4& M\\
     9 &    1277 & Star forming &    13.14 &      2.2 &     46.5$\pm$    13.6 &      6.8$\pm$     0.4& I\\
     10 &    1431 & Star forming &    27.37 &      2.9 &    212.1$\pm$    13.6 &     16.7$\pm$     0.4& (I),M \\
     11 &    1237 & Star forming &     5.49 &      2.8 &      4.5$\pm$    13.7 &      4.1$\pm$     0.4& \\
     12 &    1319 & Star forming &    55.40 &      2.4 &     47.0$\pm$    14.1 &     41.5$\pm$     0.4& \\
     13 &     723 & Star forming &    16.90 &      1.9 &    411.5$\pm$    13.7 &     51.6$\pm$     0.4& \\
     14 &     847 & Star forming &    33.12 &      2.2 &    275.7$\pm$    22.9 &     20.0$\pm$     0.4& \\
     15 &     852 & Star forming &     8.25 &      2.2 &    300.1$\pm$    18.1 &     11.2$\pm$     0.4& \\
     16 &     926 & Star forming &    20.87 &      2.0 &    316.2$\pm$    23.9 &     40.7$\pm$     0.4& \\
     17 &    1383 & Star forming &    48.50 &      2.4 &    240.9$\pm$    17.4 &     42.8$\pm$     0.4& \\
     18 &      10 & Star forming &     3.77 &      2.9 &     27.6$\pm$    13.5 &    & \\
     19 &       6 & Star forming &     8.72 &      2.0 &     80.0$\pm$    14.6 &    & \\
     20 &       7 & Quiescent &    14.09 &      1.9 &     20.4$\pm$    15.2 &     & \\
     21 &      64 & Quiescent &     9.63 &      2.4 &      -13.3$\pm$    13.9 &      3.5$\pm$     0.3& I \\
     22 &     140 & Star forming &    13.83 &      2.7 &     25.1$\pm$    14.4 &     17.2$\pm$     0.3& \\
     23 &      72 & Star forming &    23.46 &      2.0 &     55.7$\pm$    16.1 &     22.4$\pm$     0.1& \\
     24 &     914 & Quiescent &    10.95 &      2.2 &      -37.1$\pm$    14.7 &     10.9$\pm$     0.4& \\
     25 &    1061 & Star forming$^{(b)}$ &    37.65 &      2.1 &      -3.7$\pm$    13.4 &     27.1$\pm$     0.4& \\
     26 &    1100 & Star forming &    21.84 &      2.9 &    262.1$\pm$    14.7 &     34.1$\pm$     0.4& C, I, M \\
     27 &     313 & Star forming &    31.36 &      2.0 &    282.4$\pm$    13.6 &     28.9$\pm$     0.4& C, (I)\\  \hline
      1 &    1654 & Star forming &     1.82 &      2.2 &      0.0$\pm$    13.5&     1.7$\pm$     0.4&\\ 
      2 &    1658 & Quiescent &     2.01 &      3.0 &     17.0$\pm$    15.1&     0.1$\pm$     0.4&\\
      3 &    1127 & Quiescent &     0.41 &      2.2 &      -14.2$\pm$    13.5&    &\\
      4 &    1276 & Star forming &     2.12 &      3.6 &    -10.6$\pm$    14.7&   &\\
      5 &    1253 & Star forming &    14.00 &      3.0 &     68.4$\pm$    15.0&    11.6$\pm$     0.4&\\
      6 &    1227 & Star forming &     2.47 &      3.6 &      7.7$\pm$    13.9&     0.9$\pm$     0.4&\\
      7 &    1358 & Star forming &     2.77 &      2.3 &     58.2$\pm$    67.0&     6.2$\pm$     0.4&\\
      8 &     835 & Star forming &     1.67 &      3.0 &     24.0$\pm$    22.1&     2.5$\pm$     0.4&\\
      9 &     773 & Star forming &     2.15 &      2.8 &    -30.2$\pm$    14.1&     0.5$\pm$     0.4&\\
     10 &    1371 & Star forming &     3.84 &      3.0 &     36.0$\pm$    13.5&     0.6$\pm$     0.4& I\\
     11 &    1372 & Star forming &     3.74 &      2.7 &     12.7$\pm$    16.4&     3.5$\pm$     0.4&\\
     12 &    1517 & Star forming &     6.90 &      3.6 &    -48.1$\pm$    14.5&     3.4$\pm$     0.4& I \\
     13 &       5 & Star forming &    13.11 &      2.0 &    259.0$\pm$    15.8&   &\\
     14 &    1131 & Star forming &     1.63 &      3.6 &      0.4$\pm$    14.6&   &\\
\enddata
\tablecomments{$^{(a))} $ C) Chandra flux, I) Irac colors, M) MIPS 24$\micron$ flux consistent with presence of AGN. $^{(b)}$ The CSF and SSP models provide equally good fits to SED, but the  $\chi^2$ of the CSF model is marginally better. The horizontal line divides the DRG and DBG samples.}
\label{modeltable}
\end{deluxetable}

%% file: tab2.tex
\begin{deluxetable}{lrrrrrrrr}
\tablecolumns{9}
\tablewidth{0pc}
\tablecaption{Properties of $z>2$ sample}
\tablehead{
\colhead{Id} & \colhead{FIRES-Id} & \colhead{${z_{ph}}^{(a)}$} & \colhead{K$^{(b)}$} & \colhead{J-K $^{(c)}$} &  
\colhead{$z_{spec}$}  & \colhead{n} &\colhead{$r_e(\arcsec)$} &\colhead{$r_e$(kpc)} 
}
\startdata
DRG-1 &	1644	&  2.20$^{+0.10}_{-0.20}$  & 20.85 &    2.43  &       &   1  &  0.24 & 1.98\\       
DRG-2 &	1712	&  1.98$^{+0.10}_{-0.04}$  & 21.15 &    2.35  &       &   1  &  0.14 & 1.18\\ 
DRG-3 &	1612	&  2.24$^{+0.08}_{-0.06}$  & 20.15 &    3.08  &       &   2  &  0.23 & 1.93\\
DRG-4 &	1734	&  1.88$^{+0.04}_{-0.04}$  & 19.49 &    2.40  & 2.697 &   1  &  0.12 & 0.92\\
DRG-5 &	1038	&  2.44$^{+0.06}_{-0.56}$  & 21.35 &    2.50  &       &   1  &  $<0.08$ & 0.61\\
DRG-6 &	1050    &  2.96$^{+0.26}_{-0.14}$  & 21.42 &    3.07  &       &   4  &  0.93 & 1.36\\
DRG-7 &	1144	&  2.20$^{+0.14}_{-0.10}$  & 20.70 &    3.15  &       &   1  &  0.10 & 0.81\\
DRG-8 &	1198    &  2.82$^{+0.12}_{-0.14}$  & 21.60 &    3.68  &       &   1  &  0.08 & 0.61\\
DRG-9 &	1277	&  2.22$^{+0.50}_{-0.16}$  & 21.29 &    3.11  &       &   4  &  0.13 & 1.05\\
DRG-10 & 1431	&  2.92$^{+0.50}_{-0.24}$   & 20.99 &    3.06  &       &   2  &  0.41 & 3.22\\
DRG-11 & 1237	&  2.82$^{+0.12}_{-0.20}$  & 21.11 &    2.66  &       &   1  &  0.14 & 1.06\\
DRG-12 & 1319	&  2.54$^{+0.08}_{-0.12}$  & 19.02 &    2.59  & 2.424 &   3  &  0.37 & 3.02\\
DRG-13 & 723	&  1.86$^{+0.08}_{-0.04}$  & 19.46 &    2.50  &       &   2  &  0.65 & 5.46\\
DRG-14 & 847	&  2.24$^{+0.56}_{-0.18}$  & 20.32 &    2.98  &       &   1  &  0.20 & 1.61\\
DRG-15 & 852	&  2.22$^{+0.74}_{-0.24}$ & 20.95 &    2.56  &       &   1  &  0.27 & 2.19\\
DRG-16 & 926	&  1.98$^{+0.04}_{-0.06}$  & 20.32 &    2.71  &       &   1  &  0.34 & 2.84\\
DRG-17 & 1383	&  2.26 $^{+0.30}_{-0.04}$ & 19.56 &    2.89  & 2.423 &   4  & 1.26 & 10.26\\
DRG-18 & 10	&  2.88$^{+0.06}_{-0.08}$  & 21.56 &    3.39  &       &   2  &  0.10 & 0.76\\
DRG-19 & 6	&  2.00 $^{+1.78}_{-0.06}$ & 20.66 &    2.47  &       &   1  &  0.55 & 4.63\\
DRG-20 & 7	&  1.86 $^{+0.04}_{-0.06}$ & 20.58 &    2.47  &       &   1  &  $<0.08$ & 0.63\\
DRG-21 & 64	&  2.42 $^{+0.30}_{-0.06}$ & 21.56 &    3.88  &       &   2  &  $<0.08$ & 0.61\\
DRG-22 & 140	&  2.84$^{+0.10}_{-0.06}$  & 19.77 &    2.69  & 2.705 &   1  &  0.09 & 0.72\\
DRG-23 & 72	&  2.02 $^{+1.32}_{-0.02}$ & 20.20 &    2.74  &       &   1  &  0.39 & 3.00\\
DRG-24 & 914	&  2.22$^{+0.14}_{-0.08}$  & 19.81 &    2.32  &       &   1  &  $<0.08$ & 0.62\\
DRG-25 & 1061	&  2.12$^{+0.12}_{-0.04}$  & 19.63 &    3.04  &       &   3  &  0.19 & 1.55\\
DRG-26 & 1100	&  2.88$^{+0.06}_{-0.08}$  & 20.65 &    2.34  &       &   1  &  1.04 & 8.09\\
DRG-27 & 313	&  2.02$^{+0.74}_{-0.02}$  & 20.18 &    2.79  &       &   3  &  0.88 & 7.36\\ \hline
DBG-1  & 1654    & 2.22$^{+0.72}_{-0.06}$  & 21.67 &    2.27  &       &   1  &  0.12 & 0.99   \\
DBG-2  & 1658 & 3.04$^{+0.60}_{-0.04}$  &    21.12  &    2.29  &   &  1  &   0.11 & 0.83   \\
DBG-3  & 1127 & 2.16$^{+0.58}_{-0.18}$  &    21.67  &    1.39  &   &  1  &   0.10 & 0.79   \\
DBG-4  & 1276 & 3.60$^{+0.00}_{-0.22}$  &    21.40  &    1.72  &   &  2  &   0.12 & 0.86   \\
DBG-5  & 1253 & 3.00$^{+0.04}_{-0.70}$  &    20.19  &    2.05  &   &  2  &   0.99 & 7.66   \\
DBG-6  & 1227 & 3.64$^{+0.12}_{-0.28}$  &    21.64  &    1.92  &   &  1  &   0.16 & 1.12   \\ 
DBG-7  & 1358 & 2.28$^{+0.10}_{-0.12}$  &    21.26  &    2.23  &   &  1  &   0.18 & 1.45   \\
DBG-8  &  835 & 3.00$^{+0.10}_{-0.04}$ &    21.66  &    1.66  &   &  2  &   0.20 & 1.52   \\
DBG-9  &  773 & 2.76$^{+0.04}_{-0.48}$  &    20.99  &    1.46  &   &  1  &   0.09 & 0.73   \\
DBG-10 & 1371 & 2.98$^{+0.04}_{-0.60}$  &    21.52  &    1.99  &   &  1  &   0.73 & 5.63   \\
DBG-11 & 1372 & 2.68$^{+0.14}_{-0.14}$  &    21.23  &    2.28  &   &  1  &   0.80 & 6.37   \\
DBG-12 & 1517 & 3.60$^{+0.02}_{-0.50}$  &    21.42  &    1.67  &   &  4  &   $<0.08$ & 0.54   \\
DBG-13 &    5 & 2.02$^{+0.06}_{-0.08}$  &    20.35  &    2.28  &   &  1  &   0.31 & 2.55   \\
DBG-14 & 1131 & 3.60$^{+0.06}_{-0.06}$  &    21.50  &    2.19  &   &  3  &   $<0.08$ & 0.54   \\
\enddata
\tablecomments{ $^{(a)}$ Photometric redshift errors are derived from the full photometric redshift probability distribution P(z) as computed using a Monte-Carlo simulation (see Rudnick et al. 2003).  P(z) can be highly non-Gaussian, due to the complex dependence of flux errors and template type on the allowable redshift range.  For this reason the $68\%$ confidence intervals for some of the sources are highly non-symmetric.  $^{(b)}$  K-band (Vega) ``total'' magnitudes, corrected for Galactic extinction. $^{(c)}$ J-K (Vega) colors derived in colour apertures, corrected for Galactic extinction. 
The horizontal line divides the DRG and DBG samples. }
\label{sizetable}
\end{deluxetable}